\documentclass[aps,onecolumn,prb,notitlepage,twocolumn,footinbib,longbibliography]{revtex4-1}

\usepackage{amsmath}
\usepackage{amsfonts}
\usepackage{amssymb}

\usepackage[toc,page]{appendix}

\usepackage{bm}

\usepackage{graphics}
\usepackage{epsfig}
\usepackage{colortbl}
\usepackage{color}
\usepackage{nicefrac}
\usepackage{mathrsfs}
\usepackage{bbm}
\usepackage{wrapfig}

\usepackage{xcolor}
\usepackage{hyperref}

\makeatletter
\def\@seccntformat#1{\@ifundefined{#1@cntformat}%
   {\csname the#1\endcsname\quad}
   {\csname #1@cntformat\endcsname}
}
\makeatother

\catcode`\@=11
\def\numberbysection{\@addtoreset{equation}{section}
        \def\theequation{\thesection.\arabic{equation}}}

\definecolor{gruen}{rgb}{0,0.625,0}
\definecolor{rot}{rgb}{0.75,0,0}

\begin{document}

\title{Possibility of a continuous phase transition in random-anisotropy magnets with a generic random-axis distribution }

\author{D. Shapoval$^{1,2}$,  M. Dudka$^{1,2,3}$, A.A. Fedorenko$^{4}$, and Yu. Holovatch$^{1,2,5}$}

\affiliation{
$^1$\mbox{Institute for Condensed Matter Physics, National Academy of Sciences of Ukraine,
1 Svientsitskii Street, UA-79011 Lviv, Ukraine} \\
$^2$ \mbox{${\mathbb L}^4$ Collaboration \& Doctoral College for the Statistical Physics of Complex Systems, Leipzig-Lorraine-Lviv-Coventry, Europe} \\
$^3$\mbox{Institute of Theoretical Physics, Faculty of Physics, University of Warsaw, Pasteura 5, 02-093 Warsaw, Poland
}
$^4$\mbox{Universit\'e de Lyon, ENS de Lyon, Universit\'e Claude Bernard, CNRS, Laboratoire de Physique,
F-69342 Lyon, France}\\
$^5$\mbox{Centre for Fluid and Complex Systems, Coventry University, Coventry,  CV1 5FB, United Kingdom}
}


\begin{abstract}

We reconsider the problem of the critical behavior of a three-dimensional $O(m)$ symmetric magnetic system in the presence of random anisotropy disorder with a generic trimodal random axis distribution. By introducing $n$ replicas to average over disorder it can be coarse-grained to a  $\phi^{4}$-theory with $m \times n$ component order parameter and five coupling constants taken in the limit of $n \to 0$. Using a field theory approach we renormalize the model to two-loop order and  calculate the $\beta$-functions  within the $\varepsilon$ expansion and directly in three dimensions.  We analyze the corresponding renormalization group flows with the help of the Pad\'e-Borel resummation technique. We show that there is no stable fixed point accessible from  physical initial conditions whose existence was argued in the previous studies. This may indicate an absence of a long-range ordered phase
in the presence of random anisotropy disorder  with a generic random axis distribution.
\end{abstract}

\maketitle

\section{Introduction}\label{I}
\setcounter{footnote}{0}

\vspace{-0.2cm}

The structural disorder is inevitably present in many magnetic systems which undergo a phase transition.
Of particular interest is its impact near the critical points, where even weak disorder can drastically
modify the scaling behavior.\cite{Pelis02,Dotsenko95,Hol02}
One can classify different types of disorder according to the symmetry it breaks.
The most common types of disorder include: (i) random bond/site disorder where  randomness couples
linearly to the local energy density, and thus, can be viewed as local critical temperature fluctuations \cite{stichcombe-83};
(ii) random field  disorder where the order
parameter is linearly coupled to a random symmetry breaking field~\cite{imry75}; and
(iii) random anisotropy disorder in systems with continuous symmetry  where the coupling of the order parameter
to disorder is bilinear \cite{Har73}.

The effect of quenched \cite{Brout59} random bond/site disorder on the critical behavior of magnetic
systems has been studied for several decades and is now relatively well understood.
In particular according to the Harris criterion~\cite{harris74} it
modifies  the critical behavior of a  $d$-dimensional system  if the correlation length
exponent $\nu$ of the  pure system satisfies the inequality $\nu<2/d$.
The corresponding critical exponents have been computed by renormalization group (RG) methods  using $\varepsilon=4-d$ expansion up to four-loop order~\cite{folk2000}, directly in three dimensions up to six-loop order~\cite{pelissetto2000}  and using a non-perturbative approach.~\cite{Tissier2002}

The Harris criterion can be generalized to the random bond/site disorder correlated in space as a power law $\sim 1/r^a$ which is proven to be relevant for $\nu< \mathrm{max}(2/a,2/d)$.\cite{weinrib83} The corresponding critical exponents have been computed using double expansion in $\varepsilon =4-d $ and $\delta = 4-a$, \cite{weinrib83,Korutcheva1984,Korutcheva1988,Honkonen1989} directly in three dimensions,~\cite{Prudnikov1999,prudnikov00} in two dimensions using a mapping to Dirac fermions~\cite{dudka2016} and numerical simulations.\cite{Ballesteros1999,Ivanenko2008} Another model with anisotropic correlated disorder in which  extended defects are strongly correlated in $\varepsilon_d$ dimensions and randomly distributed over the remaining $d-\varepsilon_d$ dimensions was proposed in Ref.~\onlinecite{dorogovtsev-80} and studied in
Refs.~\onlinecite{boyanovsky-82,prudnikov-83,lawrie-84,yamazaki-86,decesare-94,korzhenevskii-94,blavatska-03,fedorenko2004,blavatska2005,vasilyev2015}.

\begin{figure}[]
\begin{center}
\includegraphics[width=70mm]{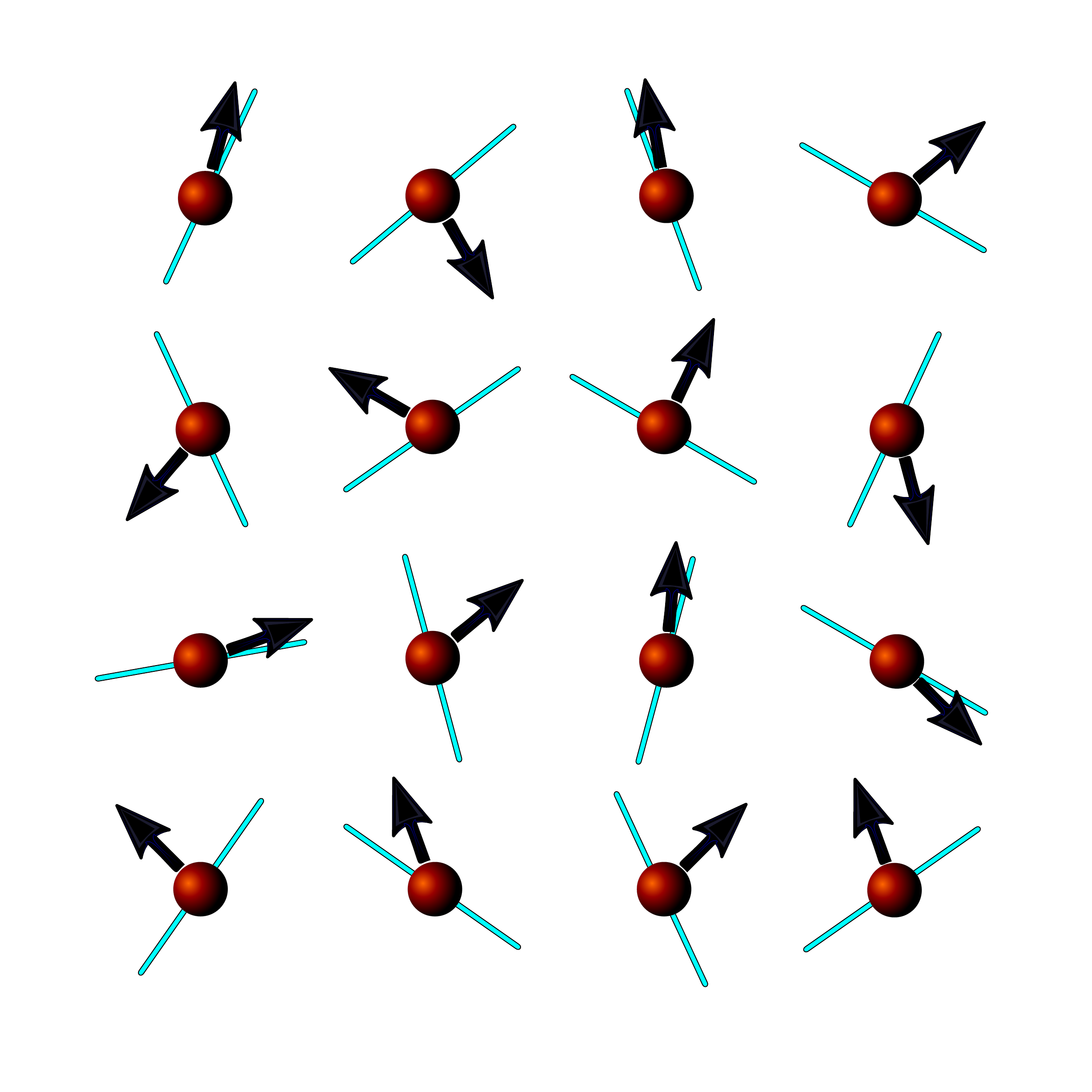}
\end{center}
\caption{  Sketch plot of the two-dimensional RAM. Red discs depict sites of the lattice with spins (black arrows) on them. Random local anisotropy axis direction on each site is shown by light blue line.  }
  \label{fig}
\end{figure}

The impact of quenched random fields and random anisotropies is usually more profound and much less studied.
For instance, a complete understanding of the simplest model, the random field Ising model (RFIM),
is still lacking despite significant numerical and analytical efforts \cite{nattermann98}.
It has been shown that the standard perturbative RG calculations lead to incorrect results due to the so-called dimensional reduction \cite{young77}.
The only known way to overcome this obstacle for the RFIM is the non-perturbative RG developed in Refs.~\onlinecite{tarjus04,balog2018}, which however, is a sophisticated and hardly controllable  method (see also recent review Ref.~\onlinecite{Tarjus2019}).
For systems with continuous symmetry the isotropically distributed random fields and random anisotropies
drive the low critical dimension of $O(m)$ symmetric systems from $d_l=2$ to $d_l=4$  with a new quasi-long-range order (QLRO)
emerging below $d_l$ \cite{feldman04}. Both the QLRO below $d_l$ and the ferromagnetic-paramagnetic transition above $d_l$ have been studied analytically using functional RG and expansion in $\varepsilon=d_l-d$ to two-loop order \cite{feldman02,doussal06,tarjus06}.
The effects of extended defects, free surfaces, and disorder correlation have been also investigated in Refs.~\onlinecite{fedorenko07,fedorenko2012,fedorenko2014,Sakamoto2019}.

The situation is even less understood in the case of an anisotropic distribution.
The critical behavior of magnets with random anisotropy is usually described by
the random anisotropy model (RAM) which  was first introduced to describe magnetic properties of amorphous alloys
of rare-earth compounds with aspherical electron distributions and transition metals \cite{Har73} (see also
Refs.~\onlinecite{Cochrane78,Sellmeyr92} for the experimental data).
The Hamiltonian of RAM can be written as
\begin{eqnarray}
\label{ham_start}
\mathcal{H} = - \sum_{<\vec{R}, \vec{R'}>} J \vec{S}_{\vec{R}} \vec{S}_{\vec{R'}} - D \sum_{\vec{R}} (\hat{x}_{\vec{R}} \vec{S}_{\vec{R}})^{2},
\end{eqnarray}
where $J > 0$ is a short-range ferromagnetic interaction between $m>1$-component spins $\vec{S}_{\vec{R}} \equiv (S_{\vec{R}}^{1}, \ldots, S_{\vec{R}}^{m})$ located on sites of a $d$-dimensional hypercubic lattice,  $\hat{x}_{\vec{R}}$ is a random unit vector indicating the direction of the local anisotropy axis on each site (see Fig.~\ref{fig}) and  $D$ is the anisotropy strength.  Here we restrict ourselves to the case of uniaxial anisotropy corresponding to $D > 0$ and do not consider an easy-plane anisotropy emergent for $D<0$.

Despite of extensive analytical and numerical  studies   even the nature of the low-temperature phase in three-dimensional random anisotropy systems is a controversial issue \cite{Goldschmidt92,Dudka05}. While for completely isotropic distribution of a random local anisotropy axis  the ferromagnetic ordering in the three-dimensional magnets is absent even in the limit of weak disorder controlled by the ratio $D/J$, it is not excluded for anisotropic distributions \cite{Dudka05}.
There is an agreement between different approaches in the case of infinitely strong disorder, where appearance of a spin-glass order was  observed \cite{Billoni2005,Mouhanna2016}. The situation is less clear for moderate and weak disorder. The question if the magnetic system  can be ordered ferromagnetically, either it will be in a QLRO or a spin-glass phase is still controversial \cite{Dudka05}.

The standard way to study the critical behavior of model (\ref{ham_start}) analytically is to coarse grain  it to a continuous effective models of $\phi^4$ type which can be averaged over disorder using replicas and  studied by field-theoretical RG methods \cite{ZinnJustin96,Amit89,Klein01}. In the case of the isotropic distribution of the random anisotropy axis this leads to a model with three distinct  $\phi^4$ terms. As was shown in
Refs.~\onlinecite{Ahar75,Dud01,Dud0001} this model has no stable physically accessible fixed point (FP)  that is in agreement with the absence of the ferromagnetic state below $d=4$ for isotropic distribution of anisotropies.
In the case of random anisotropy with the cubic distribution,  vectors $x_{\vec{R}}$ are aligned along the edges of a $m$-dimensional hypercube and the effective Hamiltonian possesses four distinct $\phi^4$ terms of different symmetries. In this case a continuous phase transition of random Ising universality class  into a ferromagnetic state was predicted below $d = 4$ \cite{Dud001,Dud0001,Cal04}.
A more general model   includes five distinct  $\phi^4$ terms \cite{Muk82}.
While this model was shown to have no stable FP to one-loop order  it was recently argued that a stable FP  appears at two-loop order in $d = 3$ \cite{Dub17}.
Here we reconsider this problem by studying the model with a generic random anisotropy disorder  to two-loop order using two different RG methods: minimal subtraction ($\rm\overline{MS}$) scheme with the $\varepsilon$ expansion and massive scheme directly in three dimension. We show that the both methods provide consistent pictures which exclude the  possibility of a continuous phase transition in this model. This indicates the absence of a long-range order in the systems with a generic random anisotropy disorder.

The paper is organized as follows. Section \ref{II} introduces the effective models for different  distributions of random anisotropy axis.
In Sec.~\ref{III} we renormalize the  generic model with a trimodal  distribution of anisotropies which includes five distinct  $\phi^4$ terms using the $\varepsilon $ expansion and directly in three dimensions to two-loop order.
In Sec.~\ref{IV} we analyze the corresponding RG flow using resummation techniques.
We summarize our results in Sec.~\ref{V}.

\vspace{-0.2cm}
\section{Effective $\phi^4$ Hamiltonians}\label{II}
\vspace{-0.2cm}

We now map the spin lattice model (\ref{ham_start}) onto  an effective
$\phi^4$ theory using the  Hubbard-Stratonovich transformation
and averaging over quenched disorder \cite{Brout59} encoded by the
local random vectors $\{\hat{x}_{\vec{R}}\}$ \cite{Ahar75,Dudka05}.
We use the replica trick \cite{Emery75} introducing  $n$ copies of the original  model
and taking the limit of $n \to 0$ at the very end.
One has to specify a particular distribution $p(\hat{x}_{\vec{R}})$  of the local random unit vectors $\hat{x}_{\vec{R}}$ in the $m$-dimensional target space. Let us consider three different cases.

In the case of the \textit{isotropic distribution} any direction of  the random unit vector $\hat{x}_{\vec{R}}$ is allowed with equal probability so that the probability distribution is given by
\begin{eqnarray}
\label{isotr_distr}
p_{i}(\hat{x}) = \frac{\Gamma(m/2)}{2 \pi^{m/2}},
\end{eqnarray}
where $\Gamma(x)$ is the Euler gamma-function.
Averaging with this distribution  leads to the effective Hamiltonian \cite{Ahar75}
\begin{eqnarray}
\label{ham_isotr}
\mathcal{H}_{\rm eff} &=& -\int d^{d} r \Big\{ \frac{1}{2} \left[\mu_{0}^{2} |\bm{\phi}|^{2}  + |\nabla \bm{\phi} |^{2}\right] + \frac{u_{0}}{4!} |\bm{\phi}|^{4}  \nonumber \\ &+& \frac{v_{0}}{4!} \sum_{\alpha=1}^{n} |\vec{\phi^{\alpha}}|^{4}+ \frac{z_{0}}{4!}\sum_{\alpha, \beta =1}^{n} \sum_{i,j =1}^{m} \phi_{i}^{\alpha} \phi_{j}^{\alpha} \phi_{i}^{\beta} \phi_{j}^{\beta}  \Big\},
\end{eqnarray}
where $\bm{\phi}=\{\vec{\phi}^\alpha (\vec{r})\}$ and $\vec{\phi}^\alpha (\vec{r}) = \{\phi_{1}^\alpha(\vec{r}), \ldots  \phi^\alpha_{m}(\vec{r})\}$ is the $n$ times replicated  $m$-component order parameter, such that $|\bm{\phi}|^{2} = \sum_{i}^m \sum_{\alpha}^n |\phi_{i}^{\alpha}|^{2}$,  and $\mu$ is the bare mass. The bare coupling constants $u_{0} > 0$, $v_{0} > 0$, $z_{0} < 0$ satisfy $z_{0}/u_{0} = -m$ (see also Table~\ref{tab00}).

In the case of the {\sl{cubic distribution}} of the local anisotropy axis the random vector $\hat{x}_{\vec{R}}$ is allowed to point along one of the $m$ axes of the hypercubic lattice with the probability distribution
\begin{eqnarray}
\label{cubic_distr}
p_{c}(\hat{x}) = \frac{1}{2m} \sum_{i=1}^{m}\Big\{\delta^{(m)}(\hat{x} - \hat{k}_{i}) + \delta^{(m)}(\hat{x} + \hat{k}_{i})\Big\},
\end{eqnarray}
where $\hat{k}_{i},\ldots,\hat{k}_{m}$ are unit vectors along the axes and $\delta(y)$ is the Dirac delta-function. Averaging over the random variables $\{\hat{x}_{\vec{R}}\}$ for the {\sl{cubic distribution}} one arrives at\cite{Ahar75}
\begin{eqnarray}
\label{ham_cubic}
\mathcal{H}_{\rm eff} &=& -\int d^{d} r \Big\{ \frac{1}{2} \left[\mu_{0}^{2} |\bm{\phi}|^{2}  + |\nabla \bm{\phi}|^{2}\right] + \frac{u_{0}}{4!} |\bm{\phi}|^{4} \nonumber \\ &+& \frac{v_{0}}{4!} \sum_{\alpha=1}^{n} |\vec{\phi^{\alpha}}|^{4} + \frac{w_{0}}{4!}\sum_{\alpha, \beta =1}^{n} \sum_{i=1}^{m} (\phi_{i}^{\alpha})^{2} (\phi_{i}^{\beta})^{2} \nonumber \\ &+& \frac{y_{0}}{4!} \sum_{i=1}^{m} \sum_{\alpha=1}^{n} (\phi_{i}^{\alpha})^{4} \Big\}.
\end{eqnarray}
Here the bare coupling constants  $u_{0}$, $v_{0}$, $w_{0}$  satisfy the conditions $u_{0} > 0$, $v_{0} > 0$, $w_{0} < 0$ and  $w_{0}/u_{0} = -m$.  Note that
the term with coupling $y_{0}$  is not present in the bare microscopic model, but it  is generated by the RG transformations so we have added it from the beginning. This, however, does not fix its sign.

Both the  isotropic distribution and  the cubic distribution can be combined into the so-called trimodal distribution\cite{Fish85, Dom93}
\begin{eqnarray}
\label{complex_distr}
p(\hat{x}) = q p_{i}(\hat{x}) + (1-q) p_{c}(\hat{x}),
\end{eqnarray}
where the direction of $\hat{x}$ is chosen either from the isotropic distribution with probability $q$  or from the cubic distribution with the probability $(1 - q)$. This leads to the effective Hamiltonian that contains all terms of the effective Hamiltonians (\ref{ham_isotr})
and (\ref{ham_cubic})\cite{Muk82,Dub17}
\begin{eqnarray}
\label{ham_anis}
\mathcal{H}_{\rm eff} &=& - \int d^{d} r \Big\{ \frac{1}{2} \left[\mu_{0}^{2} |\bm \phi|^{2}  + |\nabla \bm \phi|^{2}\right] + \frac{u_{0}}{4!} |\bm \phi|^{4} \nonumber \\ &+&  \frac{v_{0}}{4!} \sum_{\alpha=1}^{n} |\vec{\phi^{\alpha}}|^{4}+ \frac{w_{0}}{4!}\sum_{\alpha, \beta =1}^{n} \sum_{i=1}^{m} (\phi_{i}^{\alpha})^{2} (\phi_{i}^{\beta})^{2}   \nonumber \\ &+& \frac{y_{0}}{4!} \sum_{i=1}^{m} \sum_{\alpha=1}^{n} (\phi_{i}^{\alpha})^{4} + \frac{z_{0}}{4!} \sum_{\alpha, \beta =1}^{n} \sum_{i,j =1}^{m} \phi_{i}^{\alpha} \phi_{j}^{\alpha} \phi_{i}^{\beta} \phi_{j}^{\beta}  \Big\}, \ \ \
\end{eqnarray}
where the bare couplings satisfy  $u_{0} > 0$, $v_{0} > 0$, $w_{0} < 0$, $z_{0} < 0$, while the sign of $y_{0}$ is arbitrary (see also Table \ref{tab00}). However ratios $z_{0}/u_{0} $ and $w_{0}/u_{0}$ resulting from the trimodal distribution (\ref{complex_distr}) are different from those for the distributions~(\ref{isotr_distr}) and (\ref{ham_cubic}),
\begin{eqnarray}
\frac{z_{0}}{u_{0}} &=& - \frac{2 q m}  {m (1-q) + 2}\label{zu},\\
\frac{w_{0}}{u_{0} }&=& -\frac{(1-q) (m+2) m}  {m (1-q) + 2} \label{wu},
\end{eqnarray}
where for (\ref{zu}) with $q = 1$ we reproduce $z_{0}/u_{0} = -m$ for the isotropic distribution, while  for (\ref{wu}) with $q = 0$ we obtain $w_{0}/u_{0} = -m$ for the cubic distribution.

It can be also shown that the effective Hamiltonian (\ref{ham_anis}) can describe  a more general local anisotropy axis distribution.  Indeed,  it can be derived  for any distribution $p(\hat{x})$ provided that it has first two non-vanishing moments
\begin{eqnarray}
M_{i j} = \int d^{m} \hat{x} p(\hat{x}) \hat{x}^{i}\hat{x}^{j},
\end{eqnarray}
\begin{eqnarray}
M_{i j k l} = \int d^{m} \hat{x} p(\hat{x}) \hat{x}^{i}\hat{x}^{j}\hat{x}^{k}\hat{x}^{l},
\end{eqnarray}
which can be expressed as\cite{Cal04}
\begin{eqnarray}
\label{momenta}
&& M_{i j} = \frac{\delta_{i j}}{m}  \\
&& M_{i j k l} = A (\delta_{i j} \delta_{k l} + \delta_{i k}\delta_{j l} + \delta_{i l} \delta_{j k}) + B \delta_{i j}\delta_{i k}\delta_{i l}.
\end{eqnarray}
Parameters $A$ and $B$ in (\ref{momenta}) are determined by the precise form of the distribution $p(\hat{x})$ and satisfy the Cauchy inequalities $A (m +2) + B \geq 1/m$ and $3A + B \geq 1/m^{2}$. Note that the effective Hamiltonian (\ref{ham_anis}) reduces to model (\ref{ham_cubic}) for $A=0$.

The effective model (\ref{ham_anis}) can be also derived by considering the system with the random single-ion cubic anisotropy given by
\begin{eqnarray}
\label{singleion}
\mathcal{H} &=& - \sum_{\vec{R}, \vec{R'}} J_{\vec{R}, \vec{R'}} \vec{S}_{\vec{R}} \vec{S}_{\vec{R'}} - D \sum_{\vec{R}} (\hat{x}_{\vec{R}} \vec{S}_{\vec{R}})^{2} \nonumber \\
&-& V \sum_{\vec{R}} \sum_{i=1}^{m} (S_{\vec{R}}^{i})^{4},
\end{eqnarray}
where $V$ is the cubic anisotropy strength. It is straightforward to show
that averaging (\ref{singleion}) over the random variables $\{\hat{x}_{\vec{R}}\}$ with {\sl{isotropic distribution}} leads to the effective Hamiltonian~(\ref{ham_anis}) with
the bare couplings $u_{0}, v_{0} > 0$, $z_{0} < 0$  and $z_{0}/u_{0}=-m$, while the sign of $y_{0}$ depends on the sign of $V$ (see Table~\ref{tab00}). Similar to  the case of the cubic distribution the coupling $w_{0}$ is not present in the bare model but it should be added, since it is generated by the RG transformations and  may be of any
sign.\cite{Muk82}

\begin{table}
\caption{\label{tab:canonsummary} The signs of the physical couplings for the  effective Hamiltonians (\ref{ham_isotr}), (\ref{ham_cubic}), and (\ref{ham_anis}). The two last lines correspond to the effective Hamiltonian obtained from the model (\ref{ham_start}) with the distribution (\ref{complex_distr})  and the model (\ref{singleion}) with the distribution  (\ref{isotr_distr}), respectively. \label{tab00}}
\begin{center}
\begin{tabular}{|c|c|c|c|c|c|c|}
\hline
Eqs. & $u_{0}$ & $v_{0}$ & $w_{0}$ & $y_{0}$ & $z_{0}$ \\
\hline
(\ref{ham_isotr}) & $> 0$ & $> 0$ & $0$ & $0$ & $< 0$ \\
\hline
(\ref{ham_cubic}) & $> 0$ & $> 0$ & $< 0$ & $\forall$ & $0$  \\
\hline
\, (\ref{ham_start}) with (\ref{complex_distr}) $\mapsto$ (\ref{ham_anis}) & $> 0$ & $> 0$ & $< 0$ & $\forall$ & $< 0$  \\
\hline
(\ref{singleion}) with (\ref{isotr_distr}) $\mapsto$ (\ref{ham_anis}) & $> 0$ & $> 0$ & $\forall$ & $\forall$ & $< 0$  \\
\hline
\end{tabular}
\end{center}
\end{table}

Let us now discuss the conditions ensuring the physical stability of models (\ref{ham_isotr}), (\ref{ham_cubic}) and (\ref{ham_anis}).  The stability analysis can be carried out along the lines of Refs.~\onlinecite{Muk82, Cal04}.  To that end we assume that the Hamiltonian has a stable minimum characterized by the homogeneous order parameter $M$. We first consider the case when the symmetry of the ordered phase is broken with respect to (i) $\phi_{i}^{\alpha} = M$. Expanding the effective Hamiltonian (\ref{ham_isotr}) around this minimum we find that the region of stability reads\cite{Muk82}
\begin{eqnarray}
\label{cond1is}
(\text{i}) \, \, v_{0} + n u_{0} + n w_{0} > 0.
\end{eqnarray}
In the case when symmetry is broken with respect to (ii) $\phi_{i}^{\alpha} = M \delta_{\alpha 1} \delta_{i 1}$ one arrives at\cite{Muk82}
\begin{eqnarray}
\label{cond2is}
(\text{ii}) \, \, v_{0} + u_{0} + w_{0} > 0.
\end{eqnarray}
If we consider that the symmetry is broken with respect to  (iii) $\phi_{i}^{\alpha} = M \delta_{i 1}$ and (iv) $\phi_{i}^{\alpha} = M \delta_{\alpha 1}$, we obtain the same conditions (\ref{cond1is}) and (\ref{cond2is}).

Repeating the same analysis for the effective Hamiltonian (\ref{ham_cubic}) we arrive at the following stability
conditions\cite{Cal04}
\begin{eqnarray}
&(\text{i})& \, \, m n u_{0} + m v_{0} + n w_{0} + y_{0} > 0, \\
&(\text{ii})& \, \,\, \, u_{0} + v_{0} + w_{0} + y_{0} > 0, \\
&(\text{iii})& \, \, n u_{0} + v_{0} + n w_{0} + y_{0} > 0, \\
&(\text{iv})& \, \, m u_{0} + m v_{0} + w_{0} + y_{0} > 0.
\end{eqnarray}
Finally,  we obtain the regions of stability of the effective Hamiltonian (\ref{ham_anis}),
\begin{eqnarray}
&(\text{i})& \, \, m n u_{0} + m v_{0} + n w_{0} + y_{0} + m n z_{0} > 0, \label{cond1}\\
&(\text{ii})& \, \,\, \, u_{0} + v_{0} + w_{0} + y_{0} + z_{0} > 0, \\
&(\text{iii})& \, \, n u_{0} + v_{0} + n w_{0} + y_{0} + n z_{0} > 0, \label{cond2}\\
&(\text{iv})& \, \, m u_{0} + m v_{0} + w_{0} + y_{0} + m z_{0} > 0.
\end{eqnarray}
As it was discussed in Ref.~\onlinecite{Muk82}  in the replica limit $n\to 0$ the only relevant stability conditions appear to be those of replica symmetric configurations. Therefore  in our case only conditions (i) and (iii) should be considered  giving for $n=0$
\begin{eqnarray}
&(\text{i})& \, \, m v_{0} + y_{0} > 0   \\ &(\text{iii})& \,\, v_{0} + y_{0} > 0.
\end{eqnarray}

Before concluding this section let us mention that the Hamiltonian (\ref{ham_anis}) is identical to
\begin{eqnarray}
\label{ham_KL}
\mathcal{H}_{\rm eff} &=& - \int d^{d} r \left\{\frac{1}{2} \left[\mu_{0}^{2} |\bm \phi|^{2}  + |\nabla \bm \phi|^{2}\right] {+} \lambda_{0} \sum_{i = 1}^{m} \sum_{\alpha = 1}^{n} (\phi_{i}^{\alpha})^{4} \right. \nonumber \\
&+& g_{0} \sum_{i = 1}^{m} \sum_{\alpha = 1}^{n} (\phi_{i}^{\alpha})^{2} \sum_{k \neq i} (\phi_{k}^{\alpha})^{2} - \tilde{u}_{0}\sum_{i = 1}^{m} \left(\varphi_{i}^{2}\right)^{2} \nonumber \\
&-& \left.  2 \tilde{v}_{0} \!\!\sum_{1 \leq i < k \leq m}\!\varphi_{i}^{2} \varphi_{k}^{2} {-} 2 \tilde{w}_{0}\!\! \sum_{1 \leq i < k \leq m}\! \left(\sum_{\alpha = 1}^{n} \phi_{i}^{\alpha} \phi_{k}^{\alpha}\right)^{2}\right\}\nonumber \\
\end{eqnarray}
with $\varphi_{i}^{2} = \sum_{\alpha}(\phi_{i}^{\alpha})^{2}$, which was introduced in  Ref.~\onlinecite{Korzh88} to study the influence of low-symmetry defects on the continuous phase transition.
Comparing this expression with (\ref{ham_anis}) one can see that recombining components $\phi_{i}^{\alpha}$ in the Hamiltonian (\ref{ham_anis})  transforms it to the Hamiltonian (\ref{ham_KL}) with the following relations between the coupling constants $\lambda_{0} = (v_{0} + y_{0})/4!$, $g_{0} = v_{0}/4!$, $\tilde{u}_{0} = - (u_{0} + w_{0} + z_{0})/4!$, $\tilde{v}_{0} = - u_{0}/4!$, and $\tilde{w}_{0} = - z_{0}/4!$ (see Appendix~\ref{sec:app0}).

\vspace{-0.2cm}
\section{Field-theory approach}\label{III}

\vspace{-0.2cm}

The field-theoretical RG approach completed by various techniques for resummation of asymptotic series \cite{ZinnJustin96,Amit89,Klein01,LeGuillou80} is generally recognized as a powerful tool to get accurate estimates of  critical exponents for systems with random bond/site disorder.\cite{Pelis02,Folk03}  It can be even applied
to frustrated systems.\cite{Del08, Del10, Delam10, Del16} Here we apply it to study the critical properties of the RAM with a generic distribution of random anisotropy axes.

The large scale behavior of the RAM with the effective Hamiltonian (\ref{ham_anis}) can be described by one-particle irreducible ({\sl{1PI}}) vertex functions which are defined as
\begin{eqnarray}
\label{startRG}
&& \delta\left(\sum_{i}^{L} p_{i} + \sum_{j}^{N} k_{j}\right) \mathring{\Gamma}^{(L,N)}(\{p\};\{k\}; \mu_{0}^{2}; \{\mathring{\lambda}\}) \nonumber  \\ && = \int^{\Lambda_{0}} d^{d}R_{1} \ldots d^{d}R_{L} d^{d}r_{1}  \ldots d^{d} r_{N}  e^{i(\sum p_{i} R_{i} + \sum k_{j} r_{j})}  \nonumber \\
&& \times \left\langle\phi^{2}(R_{1}) \ldots \phi^{2}(R_{L}) \phi(r_{1}) \cdots \phi(r_{N}) \right\rangle_{\rm 1PI}^{\mathcal{H}_{\rm eff}} ,
\end{eqnarray}
where $\{\mathring{\lambda}\} = \{u_{0}, v_{0}, w_{0}, y_{0}, z_{0}\}$ are bare coupling constants, $\{p\}$, $\{k\}$ are external momenta, $\Lambda_{0}$ is a cut-off parameter, and $\mu_{0}$ is a bare mass. In what follows we
use the upper circle to denote the bare quantity.

In general the vertex functions~(\ref{startRG}) have a complicated tensor structure. As an example consider the vertex function ${{\mathring{\Gamma}}^{(0,4)}}{}^{i j k l}_{\alpha \beta \gamma \tau}$ which we will need to renormalize the theory. It is convenient to split it into the parts which possess the tensorial structure of the different terms in the bare model~(\ref{ham_anis}). This leads to
\begin{eqnarray}
{{\mathring{\Gamma}}^{(0,4)}}{}^{i j k l}_{\alpha \beta \gamma \tau} &=& {\mathring{\Gamma}}_{u}^{(0,4)} S_{i j k l}^{\alpha \beta \gamma \tau} {+} {\mathring{\Gamma}}_{v}^{(0,4)} S_{i j k l} F_{\alpha \beta \gamma \tau}  \nonumber \\ &+& {\mathring{\Gamma}}_{w}^{(0,4)} F_{i j k l} S_{\alpha \beta \gamma \tau}  + {\mathring{\Gamma}}_{y}^{(0,4)} F_{i j k l} F_{\alpha \beta \gamma \tau} \nonumber \\ &+& {\mathring{\Gamma}}_{z}^{(0,4)} A_{i j k l}^{\alpha \beta \gamma \tau} ,
\end{eqnarray}
where we have introduced the  tensors
\begin{eqnarray}
\label{symt}
&&F_{i j k l} = \delta_{i j} \delta_{i k} \delta_{i l}, \nonumber \\
&&S_{i j k l} = \frac{1}{3} (\delta_{i j} \delta_{k l} + \delta_{i k} \delta_{j l} + \delta_{i l} \delta_{j k}), \nonumber \\
&&S_{i j k l}^{\alpha \beta \gamma \tau} = \frac{1}{3} (\delta_{i j} \delta_{k l} \delta_{\alpha \beta} \delta_{\gamma \tau} + \delta_{i k} \delta_{j l} \delta_{\alpha \gamma} \delta_{\beta \delta} + \delta_{i l} \delta_{j k} \delta_{\alpha \tau} \delta_{\beta \gamma}), \nonumber \\
&&A_{i j k l}^{\alpha \beta \gamma \tau} = \frac{3}{2} S_{i j k l} S_{\alpha \beta \gamma \tau} - \frac{1}{2}  S_{i j k l}^{\alpha \beta \gamma \tau}, \nonumber
\end{eqnarray}
and $\delta_{a b}$ is the Kronecker symbol.

\vspace{-0.2cm}
\subsection{Renormalization}\label{III1}

\vspace{-0.2cm}

The functions (\ref{startRG}) are divergent in the limit $\Lambda_{0} \to \infty$ and have to be renormalized after a proper regularization.
We apply two different renormalization schemes,
the {\sl{massive scheme}} \cite{Parisi} and the {\sl{$\rm\overline{MS}$ scheme}}.\cite{Hooft72}  To render
the vertex functions finite we introduce
the renormalization factors  $Z_{\phi}$ for  the field $\phi$, $Z_{\phi^{2}}$ for the  $\phi^{2}$-insertion, and $Z_{\lambda_{i}}$ for the coupling constants $\lambda_{i} = u, v , w, y , z$. The bare and renormalized vertex functions are related by
\begin{eqnarray}
{\Gamma}^{(L,N)}(\{p\};\{k\}; \{{\lambda}\}) =  Z_{\phi}^{N/2} Z_{\phi^{2}}^L
\mathring{\Gamma}^{(L,N)}(\{p\};\{k\};  \{\mathring{\lambda}\}). \nonumber \\
\end{eqnarray}
The renormalization schemes differ by the normalization conditions. In the massive scheme these conditions are formulated at zero external momenta and non-zero mass, and have the following form
\begin{subequations}
\label{massscheme1}
\begin{eqnarray}
\Gamma^{(0,2)}(k; -k; \mu^{2}; \{\lambda_{i}\})\Big{|}_{k=0} &=& \mu^{2},  \\
\frac{d}{d k^{2}}\Gamma^{(0,2)}(k; -k; \mu^{2}; \{\lambda_{i}\})\Big{|}_{k=0} &=& 1, \\
\Gamma^{(0,4)}_{\lambda_{i}}(\{k\};\mu^{2}; \{\lambda_{i}\})\Big{|}_{\{k\}=0} &=& \mu^{4-d} \lambda_{i}, \ \ \ \ \ \ \\
\Gamma^{(1,2)}(p; k_1, k_2; \mu^{2}; \{\lambda_{i}\})\Big{|}_{k_1=k_2=p=0} &=& 1.
\end{eqnarray}
\end{subequations}
The renormalization factors $Z_{\lambda_{i}}$ relate the bare couplings $\mathring{\lambda}_{i}$ to the renormalized ones:
\begin{eqnarray}
\label{massscheme3}
\mathring{\lambda}_{i} = \mu^{4-d} \frac{Z_{\lambda_{i}}}{Z_{\phi}^{2}} \lambda_{i}.
\end{eqnarray}

The normalization conditions for the $\rm\overline{MS}$ scheme are fixed at zero mass and given by
\begin{subequations}
\label{minsubtr1}
\begin{eqnarray}
&&\Gamma^{(0,2)}(k, -k; \tilde{\mu}; \{\lambda_{i}\})\Big{|}_{k=0} = 0,  \\
&& \frac{\partial}{\partial k^{2}}\Gamma^{(0,2)}(k, -k; \tilde{\mu}; \{\lambda_{i}\})\Big{|}_{k^{2}=\tilde{\mu}^{2}} = 1,  \\
&&\Gamma_{\lambda_{i}}^{(0,4)}(\{k\}; \tilde{\mu}; \{\lambda_{i}\}) \Big{|}_{k_{i} k_{j} = \frac{\tilde{\mu}^{2}}{3}(4\delta_{ij}-1)} = \tilde{\mu}^{4-d} \lambda_{i},  \\
&& \Gamma^{(1,2)}(p; k, -k; \tilde{\mu}; \{\lambda_{i}\}) \Big{|}_{p^{2} = k^{2} = \tilde{\mu}^{2}, \, \, p k =- 1/3 \, \tilde{\mu}^{2}} = 1, \ \ \ \ \ \
\end{eqnarray}
\end{subequations}
where the renormalized couplings $\lambda_{i}$ are
\begin{eqnarray}
\label{minsubtr2}
\mathring{\lambda}_{i} = \tilde{\mu}^{4-d} \frac{Z_{\lambda_{i}}}{Z_{\phi}^{2}} \lambda_{i},
\end{eqnarray}
and $\tilde{\mu}$ is the external momentum scale parameter.

We now introduce the RG functions
\begin{eqnarray}
\beta_{\lambda_{i}} = \frac{\partial \lambda_{i}}{\partial \ln \bar{\mu}}, \, \, \, \, \gamma_{\phi} = \frac{\partial Z_{\phi}}{\partial \ln \bar{\mu}}, \, \, \, \, \overline{\gamma}_{\phi^{2}} = - \frac{\partial \overline{Z}_{\phi^{2}}}{\partial \ln \bar{\mu}}, \nonumber
\end{eqnarray}
where $\overline{Z}_{\phi^{2}} = Z_{\phi^{2}} Z_{\phi}$  and the derivatives are taken at fixed bare parameters. Here $\bar{\mu}$ is the renormalized mass $\mu$ in the massive scheme and the scale parameter $\tilde{\mu}$ in the $\rm\overline{MS}$ scheme.
The $\beta$- and $\gamma$-functions characterize the change of the vertex functions  under the RG transformation, and thus, allow one to calculate the scaling behavior in the critical region controlled by a FP
\begin{eqnarray}
\label{fixedEq}
\beta_{\lambda_{i}}(\{\lambda^{*}_{i}\}) = 0, \, \, i=1,\ldots,5.
\end{eqnarray}
The FP solution $\{\lambda^{*}_{i}\}$ of Eqs.~(\ref{fixedEq}) describes  the critical point of the system if it is stable and accessible from initial conditions. The FP is stable if
all the eigenvalues $\{\omega_{i}\}$ of the stability matrix
\begin{eqnarray}
\label{stabmatrix}
B_{i j} = \left.\frac{\partial \beta_{\lambda_{i}}}{\partial \lambda_{i}}\right|_{\lambda_{i}=\lambda^{*}_{i}},
\end{eqnarray}
have  positive real parts ($\mathrm{Re} \, \omega_{i} > 0$).

\vspace{-0.5cm}
\subsection{RG functions}\label{III2}

Applying the renormalization schemes (\ref{massscheme1}) -- (\ref{massscheme3}), and
(\ref{minsubtr1}) -- (\ref{minsubtr2}) to the model (\ref{ham_anis}) we obtain the RG functions to two-loop order. Introducing $\varepsilon = 4-d$ the resulting $\beta$-function calculated within the both schemes can be written in the same form
\begin{eqnarray}
\label{eqbeta}
\beta_{\lambda_{i}} = -\lambda_{i}(\varepsilon + \gamma_{\lambda_{i}} - 2 \gamma_{\phi}),
\end{eqnarray}
 once  the one-loop integral $D_2=\int \frac{d^d p}{(p^2+1)^2}$ appearing in the massive scheme is included in the redefinition of the coupling constants as $\lambda_{i} \to \lambda_{i}/D_2$, $\beta_{\lambda_{i}} \to \beta_{\lambda_{i}}/ D_2$.
The corresponding $\gamma$-functions are given by
\begin{widetext}
\begin{subequations}
\label{gammaALL}
\begin{eqnarray}
\label{gammau}
u \, \gamma_{u}
&=& -\frac{1}{6} \Big[(m n + 8) u^{2} + 2 v w + 2 v z + 2 w z + 3 z^{2} + 2 (m+2) u v + 2 (n +2) u w  + 6 u y + 2 (m + n + 1) u z\Big] \mathcal{E} \nonumber \\
&-& \frac{1}{9} \Big[2 (5 mn + 22) u^{3} + 4 v^{2} w + 4 v w^{2} + 4 v^{2} z + 16 v w z  + 4 w^{2} z + 2 (m + 8) v z^{2} + 2 (n+8) w z^{2} + 3 y z^{2} \nonumber \\
&+& 3 (m+n+3)z^{3} + 24 (m+2)u^{2}v + 24 (n+2) u^{2} w + 72 u^{2} y + 24 (m+n+1) u^{2} z + 6 (m+2) u v ^{2}\nonumber \\
&+&   6 (n + 2) u w^{2} + 36 u w y + 18 u y^{2} + 12 (n +4) u w z + 36 u y z + 3 (m n + m + n + 15) u z^{2} + 60 u v w \nonumber \\
&+& 36 u v y+ 12 (m+4) u v z \Big] \mathcal{I},
\end{eqnarray}
\begin{eqnarray}
\label{gammav}
v \, \gamma_{v} &=& -\frac{1}{6} \Big[ (m+8)v^{2} + 12 u v + 4 v w + 6 v y + 2 (m+5) v z + 6 y z \Big]
\mathcal{E}
- \frac{1}{9}  \Big[2 (5m + 22)v^{3} + 6 (m n +14) v u^{2} \nonumber \\
 &+& 2 (n+6)v w^{2} + 36 v w y + 18 v y^{2} + 24 (m+5) u v^{2} + 12 (n+6) u v w + 108 u v y + 68 v^{2} w + 72 v^{2} y \nonumber \\ &+& 12 (3 m + n +11) u v z + 4 (7 m + 29) v^{2} z + 4 (n +20) v w z + 72 u y z + 132 v y z + 24 w y z + 18 y^{2} z \nonumber \\ &+& [ (m+5)n + 17 m + 67] v z^{2} + 3 (n +14) y z^{2}\Big] \mathcal{I},
\end{eqnarray}
\begin{eqnarray}
\label{gammaw}
w \, \gamma_{w} &=& -\frac{1}{6} \Big[ (n+8)w^{2} + 12 u w + 4 v w + 6 w y + 2 (n+5) w z + 6 y z \Big]\mathcal{E} - \frac{1}{9} \Big[2 (5n + 22)w^{3} + 6 (m n +14) w u^{2} \nonumber \\ &+&  2 (m+6)w v^{2} + 36 v w y + 18 w y^{2} + 24 (n+5) u w^{2} + 12 (m+6) u v w + 108 u w y + 68 w^{2} v + 72 w^{2} y \nonumber \\ &+& 12 (m + 3 n +11) u w z + 4 (7 n + 29) w^{2} z + 4 (m +20) v w z + 72 u y z + 132 w y z + 24 v y z + 18 y^{2} z \nonumber \\ &+& [(n+5)m + 17 n + 67] w z^{2} + 3 (m +14) y z^{2}\Big] \mathcal{I},
\end{eqnarray}
\begin{eqnarray}
\label{gammay}
y \, \gamma_{y} &=& -\frac{1}{6} \Big[9 y^{2} + 8 v w + 12 v y + 12 u y + 12 w y+ 6 y z \Big] \mathcal{E} - \frac{1}{9} \Big[54 y^{3} + 96 u v w + 4 (m+18) v^{2} w + 252 v w y \nonumber \\ &+& 4 (n +18) v w^{2} + 6 (m n +14) u^{2} y + 6 (m +14) v^{2} y + 6 (n+14) w^{2} y + 12 (m + 14) u v y + 144 u y^{2} \nonumber \\ &+& 12 (n +14) u w y + 144 v y^{2} + 144 w y^{2} + 8 (m + n + 10) v w z + 12 ( m + n + 7) u y z + 126 y^{2} z \nonumber \\ &+& 12 (n + 12) w y z + 12 (m + 12) v y z + 3 (m + n + 13) y z^{2} \Big] \mathcal{I},
\end{eqnarray}
\begin{eqnarray}
\label{gammaz}
z \, \gamma_{z} &=& -\frac{1}{6} \Big[(m + n + 4)z^{2} + 12 u z + 4 v z + 4 w z  \Big]\mathcal{E} - \frac{1}{9} \Big[ (2 m n + 5 m + 5 n + 27) z^{3} + 6 (m n +14)u^{2} z\nonumber \\
&+& 2 (m +6) v^{2} z + 2 (n+6) w^{2} z
+12 w y z + 2 (5 n + 22) w z^{2} + 24 y z^{2} +  44 v w z + 12 v y z \nonumber \\
&+&    2 (5 m + 22) v z^{2}
+ 12 (m +6) u v z + 12 (n+6) u w z + 36 u y z + 12 (2 m + 2n + 15) u z^{2}  \Big] \mathcal{I},
\end{eqnarray}
\begin{eqnarray}
\label{gammap}
\gamma_{\phi} &=& - \frac{1}{9}  \Big[(m n + 2) u^{2} + (m +2) v^{2} + (n +2) w^{2} + 3 y^{2} + \frac{ m n + m + n + 3}{2} z^{2} + 2 (m + 2) u v  \nonumber \\
&+&   2 (n + 2) u w + 6 u y+ 2 (m + n + 1) u z + 6 v w+ 6 v y+ 2(m+2) v z + 6 w y
+ 2 (n+2) w z + 6 y z \Big] \,\mathcal{J},
\end{eqnarray}
\begin{eqnarray}
\label{gammap2}
\overline{\gamma}_{\phi^{2}} &=& \frac{1}{6} \Big[ (m n +2) u + (m +2) v + (n +2) w + 3 y + (m + n +1)z \Big] \mathcal{E} \nonumber \\ &-& \frac{1}{3} \Big[(m n + 2) u^{2} + (m +2) v^{2} + (n +2) w^{2} + 3 y^{2} + \frac{ m n + m + n + 3}{2} z^{2} + 2 (m + 2) u v \nonumber \\ &+& 2 (n + 2) u w + 6 u y+ 2 (m + n + 1) u z + 6 v w+ 6 v y+ 2(m+2) v z + 6 w y + 2 (n+2) w z + 6 y z \Big] \mathcal{I}.
\end{eqnarray}
\end{subequations}
\end{widetext}
The $\gamma$-functions (\ref{gammaALL}) differ for the two renormalization schemes only by values of  $\mathcal{E}$, $\mathcal{I}$, and $\mathcal{J}$.
For the  $\rm\overline{MS}$ scheme one gets $\mathcal{E} = 1$, $\mathcal{I} = 1/4$, and $\mathcal{J} = - 1/8$, while in the massive scheme $\mathcal{E} = \varepsilon$, $\mathcal{I} = \varepsilon (i_{1} - 1/2)$, and $\mathcal{J} = \varepsilon i_{2}$. Here  $i_{1}$ and $i_{2}$ are loop integrals which have to be computed in fixed dimension. In $d=3$ they are given by $i_{1} = 1/6$ and $i_{2} = - 2/27$,\cite{Nick77} while their values for general $d$ can be found in Ref.~\onlinecite{Hol_int}.

While our main goal is to analyze the above RG functions in the replica limit of $n=0$,
corresponding to a disordered system with a generic random anisotropy distribution, it is also instructive
to consider the model for arbitrary values of $m$ and $n$.
Before do that, let us check that the RG functions (\ref{eqbeta})-(\ref{gammaALL}) satisfy the properties that follow from the original model (\ref{ham_anis}) and reproduce properly the results known for reduced models. \cite{Cal04} These functions are expected to

\begin{itemize}
  \item remain invariant under the simultaneous exchange $v \leftrightarrow w$ and $m \leftrightarrow n$;

  \item reproduce the RG functions of the $m n$ model in the limit of $w = y = z = 0$ or $v = y = z = 0$ with $n \leftrightarrow m$ (see Refs.\onlinecite{Aharony,Dudka,tet_boot} and references therein);

  \item reproduce the RG functions of the ($m \times n$)-component cubic model in the limit of $v= w = z=0$
     ( see Refs.~\onlinecite{Aharony, Pelis02,  Sokolov, cub_boot} and references therein);

  \item reproduce the RG functions of the randomly dilute cubic model for $w = z = 0$ and $n = 0$,\cite{Cal03} and of the tetragonal model for $w = z = 0$ with $m = 2$;\cite{Pelis02,tet_boot}

  \item satisfy for $v  = z = 0$,  and $n = 0$ the identities
\begin{eqnarray}
\beta_{u}\!(u, 0, w, y, 0) {+} \beta_{w}\!(u, 0, w, y, 0)&{=}& \beta_{RIM, u}\!(u {+} w, y),\nonumber \\
\beta_{y}\!(u, 0, w, y, 0) &{=}& \beta_{RIM, y}\!(u {+} w, y),
\end{eqnarray}
    where $\beta_{RIM, u}(u, y)$ and $\beta_{RIM, y}(u, y)$ are the RG functions of the random Ising model (RIM);\cite{Folk03}

  \item reproduce for $z = 0$ and $n=0$ the RG functions of the RAM with cubic distribution  obtained in Refs.~\onlinecite{Dud0001, Dudka05, Cal04};

  \item  reproduce for  $w = y = 0$ and  $n = 0$ the RG functions of the RAM with isotropic  distribution obtained in Ref.~\onlinecite{Dud01, Dudka05};

  \item reproduce for $n=0$ (after applying the transformation described in Appendix \ref{sec:app0})  the $\rm\overline{MS}$ $\beta$-functions
  for the crystal with low-symmetry defects derived in Ref.~\onlinecite{Korzh88}.
\end{itemize}

We have checked that our $\beta$-functions satisfy all these properties.
Note, that the two-loop $\beta$-functions  derived  in Ref.~\onlinecite{Dub17} using a massive RG scheme  do not satisfy all these conditions. For instance, the first property from the list above does not hold. As  functions of  Ref.~\onlinecite{Dub17} have been presented only for $n= 0$, to check this property we set $m = 0$ in $\beta_{v}$ of Ref.~\onlinecite{Dub17}  and substitute  $v \leftrightarrow w$. Then we  compare this with $\beta_{w}$ of Ref.~\onlinecite{Dub17} where we also set $m = 0$. The obtained functions do not coincide, as they should. Moreover for $z=0$ the RG-functions obtained in Ref.~\onlinecite{Dub17} do not match completely with (\ref{eqbeta})-(\ref{gammaALL}) and with the functions derived in Ref.~\onlinecite{Dud001}. They also do not reproduce  the RG-functions  calculated for the RAM with isotropic  distribution of anisotropies in Ref.~\onlinecite{Dud01}.

\vspace{-0.2cm}
\section{RG analysis}\label{IV}
\vspace{-0.2cm}

We can analyze the two-loop beta functions  (\ref{eqbeta})-(\ref{gammaALL}) either developing the $\varepsilon$-expansion, or  directly in $d = 3$  by setting $\varepsilon=1$  and considering the renormalized couplings  as the expansion parameters.\cite{Schloms} Since in the last case the series in the coupling constants are asymptotic, in order to get reliable numerical data one has to apply appropriate resummation techniques.\cite{ZinnJustin96,Amit89,Klein01}  In the next two subsections we will use both these approaches: we analyze our functions in the one-loop approximation using
$\varepsilon$-expansion and than apply a resummation technique to the  two-loop expressions in fixed space dimensions $d=3$.

\vspace{-0.2cm}
\subsection{One-loop approximation}\label{III3}
\vspace{-0.2cm}

Although our main interest is to analyze the RG-functions (\ref{eqbeta})-(\ref{gammaALL}) in the limit of $n = 0$, the model under consideration has some applications also for non-zero $n$. The simplest example is the $mn$-vector model.\cite{Aharony, Dudka}. At $m = 1$ and arbitrary $n$ it reduces to the cubic model, \cite{Aharony,Pelis02}  while for $m = 2$, $n = 2$ and $n = 3$ it describes a class of special structural phase transitions.\cite{Muk76} Another example is provided by the systems described by the reduced effective Hamiltonian (\ref{ham_anis}) with $w=z=0$. At  $n = 0$ it  corresponds to the  randomly dilute cubic model\cite{Cal03} and for   $m = 2$ and non-zero $n$ it corresponds to the tetragonal model.\cite{Pelis02}

\begin{table*}
\caption{FPs as a function of  $m$ and $n$ to the first order in $\varepsilon$. Only  $22$ FPs (from all $32$ FPs), which can be calculated analytically, are shown. The rest $10$ FPs  are discussed in  Appendix \ref{sec:secondapp}. Here, $x_{\pm} = (m + n - 2 \pm \sqrt{(m + n - 2)^{2} - 12 m n + 48})/ (8 - 2m n)$; $A_{\pm} (m, n) = (m + n - 2 + 2 m \sigma(m, n) \pm \sqrt{(m + n - 2 + 2 m \sigma(m, n))^{2} + 4 (4 - m n)(2 \sigma(m, n) + 3)})/ (8 - 2m n)$, $\sigma(m, n) = - (m - n + 6)/(m + 4)$; $A_{\pm}(n, m) = (m + n - 2 + 2 n \sigma(n, m) \pm \sqrt{(m + n - 2 + 2 n \sigma(n, m))^{2} + 4 (4 - m n)(2 \sigma(n, m) + 3)})/ (8 - 2m n)$,  $\sigma(n, m) = - (n - m + 6)/(n + 4)$; $\alpha_{\pm} = ((n-4)\gamma + 2 m \pm \sqrt{((n-4)\gamma + 2 m)^{2} + 8 (4 - m n)\gamma})/(8 - 2 m n)$, $\beta_{\pm} = - ((4 - n)\gamma + 8 \pm \sqrt{((4 - n)\gamma + 8)^{2} - 96\gamma})/6$, $\gamma = (m + 4)/(n + 4)$, $B_{\pm \pm} = 12 \alpha_{\pm} + 6 \beta_{\pm} + (n + 8)\gamma +4$, $\rho = m + n + 4$, $\zeta(m, n) = (m n + 8) (m + 8)$ , $\Sigma_{\pm}(m, n) = \sigma(m, n)+3A_{\pm}(m, n)$. \label{tab0}}
\begin{center}
\begin{tabular}{|c|c|c|c|c|c|}
\hline
FP & $u^{*}$ & $v^{*}$ & $w^{*}$ & $y^{*}$ & $z^{*}$ \\
\hline
I. & 0 & $ 0 $ & $0$ & $0$ & $0$ \\
\hline
II. & 0 & $ \frac{6}{m + 8}\varepsilon $ & $0$ & $0$ & $0$ \\
\hline
III. & $\frac{6}{m n + 8}\varepsilon$ & $ 0 $ & $0$ & $0$ & $0$  \\
\hline
IV. & 0 & $ 0 $ & $\frac{6}{n + 8}\varepsilon$ & $0$ & $0$  \\
\hline
V. & 0 & $ 0 $ & $0$ & $\frac{2}{3}\varepsilon$ & $0$  \\
\hline
VI. & $\frac{6 (m - 4)}{24 (m + 2) - \zeta(m, n)}\varepsilon$ & $\frac{6 (m n - 4)}{\zeta(m, n) - 24 (m + 2)}\varepsilon$ & $0$ & $0$ & $0$  \\
\hline
VII. & $\frac{6 (n - 4)}{24 (n + 2) - \zeta(n, m)}\varepsilon$ & $ 0 $ & $\frac{6 (m n - 4)}{\zeta(n, m) - 24 (n + 2)}\varepsilon$ & $0$ & $0$  \\
\hline
VIII. & 0 & $ \frac{2}{m} \varepsilon$ & $0$ & $\frac{2 (m - 4)}{3 m} \varepsilon$ & $0$  \\
\hline
IX. & $\frac{2 (m - 4)}{(8 - m n)m - 16}\varepsilon$ & $\frac{2 (4 - m n)}{(8 - m n)m - 16}\varepsilon$ & $0$ & $\frac{2}{3}\frac{(m n - 4) (m - 4)}{(m n - 8)m + 16}\varepsilon$ & $0$  \\
\hline
X. & $\frac{2 (n - 4)}{(8 - m n)n - 16}\varepsilon$ & $ 0 $ & $\frac{2 (4 - m n)}{(8 - m n)n - 16}\varepsilon$ & $\frac{2}{3}\frac{(m n - 4) (n - 4)}{(m n - 8)n + 16}\varepsilon$ & $0$  \\
\hline
XI. $a$ & $\frac{6 \alpha_{+}}{B_{++}}\varepsilon$ & $\frac{6}{B_{++}}\varepsilon$ & $\frac{6 \gamma}{B_{++}}\varepsilon$ & $\frac{6 \beta_{+}}{B_{++}}\varepsilon$ & $0$\\
\hline
\qquad $b$ & $\frac{6 \alpha_{+}}{B_{+-}}\varepsilon$ & $\frac{6}{B_{+-}}\varepsilon$ & $\frac{6 \gamma}{B_{+-}}\varepsilon$ & $\frac{6 \beta_{-}}{B_{+-}}\varepsilon$ & $0$\\
\hline
\qquad $c$ & $\frac{6 \alpha_{-}}{B_{-+}}\varepsilon$ & $\frac{6}{B_{-+}}\varepsilon$ & $\frac{6 \gamma}{B_{-+}}\varepsilon$ & $\frac{6 \beta_{+}}{B_{-+}}\varepsilon$ & $0$\\
\hline
\qquad $d$ & $\frac{6 \alpha_{-}}{B_{--}}\varepsilon$ & $\frac{6}{B_{--}}\varepsilon$ & $\frac{6 \gamma}{B_{--}}\varepsilon$ & $\frac{6 \beta_{-}}{B_{--}}\varepsilon$ & $0$\\
\hline
XII. & $\frac{2}{m n} \varepsilon$ & $0$ & $0$ & $\frac{2}{3} \frac{m n - 4}{m n} \varepsilon$ & $0$  \\
\hline
XIII. & $0$ & $0$ & $\frac{2}{n} \varepsilon$ & $\frac{2}{3} \frac{n - 4}{n} \varepsilon$ & $0$  \\
\hline
XIV. $a$ & $\frac{6 x_{+}}{\rho + 12 x_{+}} \varepsilon$ & $0$ & $0$ & $0$ & $\frac{6}{\rho + 12 x_{+}} \varepsilon$ \\
\hline
\qquad $b$ & $\frac{6 x_{-}}{\rho + 12 x_{-}} \varepsilon$& $0$ & $0$ & $0$ & $\frac{6}{\rho + 12 x_{-}} \varepsilon$  \\
\hline
XV. $a$ & $\frac{6 A_{+}(m, n) \varepsilon}{\rho + 4 \Sigma_{+}(m, n)}$& $\frac{6 \sigma(m, n) \varepsilon}{\rho + 4 \Sigma_{+}(m, n)}$ & $0$ & $0$ & $\frac{6 \varepsilon}{\rho + 4 \Sigma_{+}(m, n)}$  \\
\hline
\qquad $b$ & $\frac{6 A_{-}(m, n) \varepsilon}{\rho + 4 \Sigma_{-}(m, n)}$& $\frac{6 \sigma(m, n) \varepsilon}{\rho + 4 \Sigma_{-}(m, n)}$ & $0$ & $0$ & $\frac{6 \varepsilon}{\rho + 4 \Sigma_{-}(m, n)}$  \\
\hline
XVI. $a$ & $\frac{6 A_{+}(n, m)\varepsilon}{\rho + 4 \Sigma_{+}(n, m)}$& $0$ & $\frac{6 \sigma(n, m) \varepsilon}{\rho + 4 \Sigma_{+}(n, m)}$ & $0$ & $\frac{6 \varepsilon}{\rho + 4 \Sigma_{+}(n, m)}$  \\
\hline
\qquad $b$ & $\frac{6 A_{-}(n, m)\varepsilon}{\rho + 4 \Sigma_{-}(n, m)}$& $0$ & $\frac{6 \sigma(n, m)\varepsilon}{\rho + 4 \Sigma_{-}(n, m)}$ & $0$ & $\frac{6 \varepsilon}{\rho + 4 \Sigma_{-}(n, m) } $  \\
\hline
\end{tabular}
\end{center}
\end{table*}

Let us first analyze the FPs of the RG functions (\ref{eqbeta})-(\ref{gammaALL}) to the first-order in $\varepsilon$ for arbitrary values of $m$ and $n$. To this order the  RG functions derived using the both schemes coincide and read
\begin{widetext}
\begin{subequations}
\label{RGu1LA}
\begin{eqnarray}
\beta_{u} &=& - \varepsilon u + \frac{1}{6} \Big[(m n + 8) u^{2} + 2 v w + 2 v z + 2 w z + 3 z^{2} + 2 (m+2) u v {+} 2 (n +2) u w {+} 6 u y + 2 (m + n + 1) u z \Big], \\
\beta_{v} &=& - \varepsilon v +\frac{1}{6} \Big[ (m+8)v^{2} + 12 u v + 4 v w + 6 v y + 2 (m+5) v z + 6 y z \Big],  \\
\beta_{w} &=& - \varepsilon w +\frac{1}{6} \Big[ (n+8)w^{2} + 12 u w + 4 v w + 6 w y + 2 (n+5) w z + 6 y z \Big],  \\
\beta_{y} &=& - \varepsilon y +\frac{1}{6} \Big[9 y^{2} {+} 8 v w {+} 12 v y {+} 12 u y + 12 w y{+} 6 y z \Big],  \\
\beta_{z} &=& - \varepsilon z +\frac{1}{6} \Big[(m + n + 4)z^2 + 12 u z + 4 v z + 4 w z  \Big].
\end{eqnarray}
\end{subequations}
\end{widetext}
The system of equations (\ref{RGu1LA}) has $32$ solutions, from which the first 16 FPs has $z = 0$, and thus, describe a system with the cubic anisotropy distribution (\ref{cubic_distr}).

They are shown in lines I -- XIII  of Table~\ref{tab0}  where we group  the FPs with the same vanishing coupling constant. The first $14$ FPs being taken in the limit of $n \to 0$ match  those found in Ref.~\onlinecite{Ahar75, Dud001, Cal04}. Note that the coordinates of several FPs have a pole at $n \to 0$ (e.g. FP XII and XIII in Table~\ref{tab0}), and thus, do not exist in this limit. The corresponding FPs with ($u^{*}\not=0$, $y^{*}\not=0$, $v^{*} = w^{*} = z^{*} = 0$ and $w^{*}\not=0$, $y^{*}\not=0$, $u^{*} = v^{*} = z^{*} = 0$) can be obtained in the next order of approximation with the help of $\sqrt{\varepsilon}$-expansion.\cite{sqrte,Folk03} This also applies to the FP IX at $n = 0$ and $m = 2$.\cite{Dud001}

The rest 16 FPs with  $z^* \neq 0$ can be found along the lines of Ref.~\onlinecite{Korzh88} (see Appendix~\ref{sec:secondapp}).   Out of them, only six can be expressed in the analytic form, the coordinates of the rest $10$ FPs can be found only numerically.
The FPs XIV -- XVI with $z^* \neq 0$ which can be computed analytically are shown in Table~\ref{tab0}.
Stability analysis of FPs listed in the Table~\ref{tab0} and other 10 FPs found numerically  at $n = 0$ does not indicate that there are other stable FPs except for the FP III (for details see Appendix~\ref{sec:secondapp}). However, as it has been pointed out in Refs.~\onlinecite{Dudka05,Dud01,Dud0001,Dud001,Cal04}
where the reduced versions of the Hamiltonian (\ref{ham_anis}) were analyzed using RG methods, this FP can not be reached along the RG flow starting from physical initial conditions. Indeed, the bare coupling constants satisfy
conditions (\ref{zu}) and  (\ref{wu}) and have fixed signs outlined below Eq.~(\ref{ham_anis}). 
The RG flow starting in this region will never reach FP III because of separatrices that restrict its basin of attraction.

We also computed the FPs for  $m = 2$, $m=3$ and  $n = 1$, $n = 2$,  $n = 3$ which are shown in  Tables~\ref{tabc1}-\ref{tabc8} of Appendix~\ref{sec:secondapp}. We find that the FP III  is also the only stable FP for $n=1$, while for $n>1$ there is no stable FP.

Therefore, the exhaustive analysis of the one-loop $\beta$ functions indicates the absence of a continuous phase transition of the random anisotropy  with a generic random axis distribution. In the next subsection we show that this conclusion holds also at the two-loop order contrary to the claim of Ref.~\onlinecite{Dub17}.

\vspace{-0.2cm}
\subsection{Two-loop approximation}\label{III5}
\vspace{-0.2cm}

As it was shown for the  model with three coupling constants the straightforward calculation of  the FP coordinates using the asymptotic series is not very accurate.\cite{Dud01}  To extract the reliable information, we apply the Pad\'e-Borel  resummation method\cite{Bak78} which is described in the Appendix \ref{sec:firstapp}.

In this subsection we analyze the $\beta$-functions (\ref{eqbeta})-(\ref{gammaALL}) in the replica limit of $n=0$ for  $m = 2$, $m = 3$. To that end we resume  them using the Pad\'e-Borel method (\ref{serBorel}) and then solve the obtained system of five non-linear equations. The computed FPs are shown in Tables~\ref{tab1},~\ref{tab2} (for the massive RG scheme) and in Tables~\ref{tab3},~\ref{tab4} (for the $\rm\overline{MS}$ scheme). There we list only the FPs with real coordinates.  In the limiting cases, the obtained results reproduce the known ones.\cite{Dud001,Dud01,Dud0001, Dudka05}

Unlike the one-loop approximation, where we know the number of solutions, here we solve the system of non-algebraic equations and thus the number of FPs is unknown in advance.  
This procedure may lead to spurious FPs which are not perturbative in $\varepsilon$, i.e. do not coincide with the Gaussian FP in $d=4$ and which appear and disappear once one increases the number of loops taken into account. If such a solution exists and  turns out to be stable, one needs a careful analysis to check if this is a real or spurious FP, see e.g. Refs. \onlinecite{Del08,Del10,Delam10}. Fortunately, we do not find such solutions, since all FPs turn out to be unstable. For the sake of convenience  we adopt the classification of one-loop FPs introduced in Table~\ref{tab0} by regrouping all the two-loop FPs of the same symmetry found using the resummation technique.

Note that in this approximation the FP coordinates are renormalization scheme dependent and differ for the massive and $\rm\overline{MS}$ schemes.\cite{Amit89}

We consider only the  physical FPs with couplings $u^{*} > 0$, $v^{*} > 0$, $z^{*} < 0$, and any $w^{*}$ and $y^{*}$ (see Table~\ref{tab0}). Among all FPs there is only one stable physical FP. This is the ``polymer'' $\mathcal{O}$($n=0$) FP, which is stable for any $m$ (point III in Tables~\ref{tab1}~--~\ref{tab4}), but unfortunately this FP is unreachable from physical initial conditions. The FP with coordinates $u^{*} = v^{*} = z^{*} =0$, $w^{*} < 0$, and $y^{*} > 0$, which corresponds to the stable FP of the Hamiltonian (\ref{ham_cubic}), has one negative stability eigenvalue associated with coupling $z$. Thus the stable and physically accessible FP  of the RAM with the cubic distribution of local anisotropy axis (\ref{cubic_distr})  (FP XIII of Tables~\ref{tab1}~--~\ref{tab4}) becomes unstable with respect to this perturbation.

Let us compute the  corresponding crossover exponent $\phi_{z}$ which is  related to the stability eigenvalue
\begin{equation}
\omega_{z} = \frac{\partial \beta_{z}}{\partial z}\Big{|}_{0,0, w^{*}, y^{*}, 0}.
\end{equation}
as $\phi_{z} = - \omega_{z} \nu$, where $\nu$ is the correlation length critical exponent calculated in this fixed point (see e.g. \onlinecite{Dud001,Dudka05}).
In the massive scheme we find
\begin{eqnarray}
\omega_{z} = - 1.1947, \qquad \phi_{z} \equiv - \omega_{z} \nu = 0.8071,
\end{eqnarray}
while in the $\rm\overline{MS}$ scheme we obtain
\begin{eqnarray}
\omega_{z} = - 1.0747, \qquad \phi_{z} \equiv - \omega_{z} \nu = 0.7173.
\end{eqnarray}
The difference between the results computed using different renormalization schemes provides an estimation of the error bars for the critical exponent values.

It is instructive to compare our result  with the six-loop estimate obtained within the massive RG scheme in Ref.~\onlinecite{Cal04}, where the RG dimension $y_{z} = - \omega_{z}$ calculated from a certain scaling operator of the cubic model is $y_{z} = 1.16(6)$, and the crossover exponent is $\phi_{z} = 0.79(4)$. Surprisingly our two-loop estimates of these universal quantities are very close to those obtained within the six-loop approximation. Such high values of crossover exponents mean that the presence of even a very small  $z$- contribution in (\ref{ham_anis}) leads to high instability of the FP XIII.

The analysis of the two-loop $\beta$ functions calculated using two different renormalization schemes
gives a solid evidence of the fact that there are no FPs that are simultaneously stable and reachable from physical initial conditions for the Hamiltonian (\ref{ham_anis}).

\begin{table*}
\caption{\label{tab:canonsummary} FPs for  $m=2$ computed in the massive scheme to two-loop order. \label{tab1}}
\begin{center}
\begin{tabular}{|c|c|c|c|c|c|}
\hline
FP  & $u^{*}$ & $v^{*}$ & $w^{*}$ & $y^{*}$ & $z^{*}$ \\
\hline
I  & 0 & $ 0 $ & $0$ & $0$ & $0$ \\
\hline
II & 0 & $ 0.9107 $ & $0$ & $0$ & $0$ \\
\hline
III & $1.1857$ & $ 0 $ & $0$ & $0$ & $0$  \\
\hline
IV & 0 & $ 0 $ & $1.1857$ & $0$ & $0$  \\
\hline
V & 0 & $ 0 $ & $0$ & $1.0339$ & $0$  \\
\hline
VI &$-0.0322$ & $0.9454$ & $0$ & $0$ & $0$  \\
\hline
VII & $2.1112$ & $ 0 $ & $-2.1112$ & $0$ & $0$  \\
\hline
VIII & 0 & $ 1.5509 $ & $0$ & $-1.0339$ & $0$  \\
\hline
IX   & $-0.4401$  & $2.3900$ & $0$ & $-1.5933$ & $0$ \\
\hline
X. $\alpha$  & ${ -0.1387}$ & $ 0 $ & ${ -0.2667}$ & ${1.5509}$ & $0$  \\
\qquad $\beta$  & $0.6678$ & $ 0 $ & $-0.6678$ & $1.0339$ & $0$  \\
\hline
XI.     $\alpha$      & ${ -0.0899}$ & ${ -0.0081}$ & ${ -0.3262}$ & ${ 1.5727}$ & $0$ \\
\qquad  $\beta$      & $0.3486$ & $-0.2398$ & $-0.4969$ & $ 1.4538$ & $0$ \\
\qquad  $\gamma$       & $0.4128$ & $0.5013$ & $0.7676$ & $ -0.5706$ & $0$ \\
\qquad  $\delta$       & ${ 0.4755}$ & ${ 1.2862}$ & ${ 1.1146}$ & ${ -2.4093}$ & $0$ \\
\qquad  $\epsilon$      & $1.9951$ & $-1.7745$ & $-2.4995$ & $ 1.9710$ & $0$ \\

\hline
XII    & $0.4755$  & $0$ & $0$ & $-2.4093$ & $0$ \\
\hline
XIII   & $0$ & $0$ & $-0.4401$ & $1.5933$ & $0$  \\

\hline		
XIV. $\alpha$     			 & $0.5349$ & $ 0 $ & $0$ & $0$ & $0.5325$  \\
\qquad \enspace  $\beta$     & $1.4650$ & $ 0 $ & $0$ & $0$ & $-1.6278$  \\
\hline
XV.  $\alpha$    & ${ 0.3427}$ & ${ 2.0830}$ & $0$ & $0$ & ${ -1.1498}$  \\
\qquad \enspace  $\beta$    & $0.7991$ & $ 0.7341 $ & $0$ & $0$ & $-0.5360$  \\
\hline
XVI. $\alpha$     & $0.5929$ & $0$ & $-1.1857$ & $0$ & $1.1857$ \\
\qquad \enspace  $\beta$      & $1.0556$ & $0$ & $2.1112$ & $0$ & $-2.1112$ \\
\hline
XVII. $\alpha$    & $-0.2201$ & $2.3900$ & $0.4401$ & $-1.5933$ & $-0.4401$  \\
\qquad \enspace $\beta$  	  & $0.1106 $ & $1.9238$ & $0.5040$ & $-1.4409$ & $-0.5040$  \\
\qquad \enspace $\gamma$    & $ 0.3339$ & $1.5509$ & $0.6678$ & $-1.0339$ & $-0.6678$  \\
\qquad \enspace $\delta$    & $0.7139$ & $1.1670$  & $2.4589$ & $-1.9465$ & $-2.4077$  \\
\qquad \enspace $\epsilon$  & $0.7394$ & $1.1381$  & $2.4016$ & $-1.8889$ & $-2.4016$  \\
\qquad \enspace $\zeta$     & $0.7971$ & $-0.3573$ & $-0.7735$& $0.5750 $ & $0.7735$  \\
\hline
\end{tabular}
\end{center}
\end{table*}

\begin{table*}
\caption{\label{tab:canonsummary} FPs for $m=3$ computed in the massive scheme to two-loop order. \label{tab2}}
\begin{center}
\begin{tabular}{|c|c|c|c|c|c|}
\hline
FP  & $u^{*}$ & $v^{*}$ & $w^{*}$ & $y^{*}$ & $z^{*}$ \\
\hline
I  & 0 & $ 0 $ & $0$ & $0$ & $0$ \\
\hline
II & 0 & $ 0.8102 $ & $0$ & $0$ & $0$  \\
\hline
III  & $1.1857$ & $ 0$ & $0$ & $0$ & $0$  \\
\hline
IV & $0$ & $ 0 $ & $1.1857$ & $0$ & $0$  \\
\hline
V  & 0 & $ 0 $ & $0$ & $1.0339$ & $0$  \\
\hline
VI &$0.1733$& $ 0.6460 $ & $0$ & $0$ & $0$  \\
\hline
VII & $2.1112$ & $ 0 $ & $-2.1112$ & $0$ & $0$  \\
\hline
VIII & $0$ & $ 0.8394 $ & $0$ & $-0.0485$ & $0$  \\
\hline
IX  & $0.1695$ & $ 0.7096 $ & $0$ & $-0.1022$ & $0$  \\
\hline
X. $\alpha$   & $0.6678$ & $ 0 $ & $-0.6678$ & $1.0339$ & $0$  \\
\qquad $\beta$    & $-0.1387$ & $ 0 $ & $-0.2667$ & $1.5509$ & $0$  \\
\hline
XI. $\alpha$      & ${ -0.0879}$ & ${ -0.0070}$ & ${ -0.3295}$ & ${ 1.5731}$ & $0$ \\
\qquad  $\beta$      & $0.2833$ & $-0.1901$ & $-0.5381$ & $1.5365$ & $0$ \\
\qquad   $\gamma$       & $0.4371$ & $0.4027$ & $0.7289$ & $-0.5051$ & $0$ \\
\qquad   $\delta$      & $0.5704$ & $1.0219$ & $1.1630$ & $-2.2717$ & $0$ \\
\hline
XII    & $0.4755$  & $0$ & $0$ & $-2.4093$ & $0$ \\
\hline
XIII   & $0$ & $ 0 $ & $-0.4401$ & $1.5933$ & $0$  \\
\hline
XIV    & $0.5386$ & $ 0 $ & $0$ & $0$ & $0.4431$  \\
\hline
XV     & $0.8450$ & $ 0.5934 $ & $0$ & $0$ & $-0.4506$  \\
\hline
XVI    & $0.5753$ & $0$ & $-0.4570$ & $0$ & $0.6462$ \\
\hline
XVII    & $0.8962$ & $-0.3497$ & $-0.8276$ & $0.7597$ & $0.5187$  \\
\hline
\end{tabular}
\end{center}
\end{table*}

\begin{table*}
\caption{\label{tab:canonsummary} FPs for  $m = 2$ computed in the $\rm\overline{MS}$ scheme to two-loop order. \label{tab3}}
\begin{center}
\begin{tabular}{|c|c|c|c|c|c|}
\hline
FP  & $u^{*}$ & $v^{*}$ & $w^{*}$ & $y^{*}$ & $z^{*}$ \\
\hline
I   & $0$ & $ 0 $ & $0$ & $0$ & $0$ \\
\hline
II  &  $0$ & $1.1415$ & 0 & 0 & 0  \\
\hline
III & $1.5281$ & 0 & 0 & 0 & 0  \\
\hline
IV  & 0 & 0 & 1.5281  & 0 & 0  \\
\hline
V   & 0 & 0 & 0 & 1.3146 & 0  \\
\hline
VI  &  0.1429 & 0.9923 & 0 & 0 & 0 \\
\hline
VII &  2.5382 & 0 & -2.5382 & 0 & 0  \\
\hline
VIII. $\alpha$           &  0 & -0.6347 & 0 & 2.1354 & 0   \\
\qquad \enspace  $\beta$ &  0 & 1.9719 & 0 & -1.1346 & 0   \\
\hline			
IX. $\alpha$     & -0.2506 & 2.4494 & 0 & -1.6330 & 0  \\
\qquad $\beta$   & -0.2273 & 0.0544 & 0 & 1.5335& 0  \\
\hline
X. $\alpha$       & -0.0328 & 0 & -0.2134 & 1.6275 & 0 \\
\qquad $\beta$    & 0.7311 & 0 & -0.7311 & 1.3146 & 0   \\
\hline
XI. $\alpha$       & -0.1940 & 0.0306  & -0.0400 & 1.5737  & 0   \\
\qquad $\beta$     & -0.0228 & -0.0003  & -0.2247 & 1.6294  & 0   \\
\qquad $\gamma$    & 0.2670  & -0.1330 & -0.4058 & 1.6247  & 0   \\
\qquad $\delta$    & 0.5580  & 0.6121  &  0.9464 & -0.6988 & 0   \\
\qquad $\epsilon$  & 0.5580  & 1.5704  &  1.2423 & -2.7081 & 0   \\
\qquad $\zeta$     & 2.3469  & -2.1042 & -2.8990 & 2.3216  & 0   \\
\hline
XII   & -0.2506 & 0 & 0 & 1.6330 & 0 \\
\hline
XIII  & 0 & 0 & -0.2506 & 1.6330 & 0  \\
\hline
XIV. $\alpha$   & 0.7060 & 0 & 0 & 0 & 0.6578   \\
\qquad \enspace  $\beta$    & 1.6637 & 0 & 0 & 0 & -1.8212  \\
\hline	
XV. $\alpha$    			& 0.4515 & 2.2913 & 0 & 0 & -1.2002  \\
\qquad \enspace  $\beta$    & 1.0126 & 0.9058 & 0 & 0 & -0.6522  \\
\hline
XVI. $\alpha$   & 0.7641 & 0 & -1.5281 & 0 & 1.5281   \\
\qquad \enspace  $\beta$    & 1.2691 & 0 &  2.5382 & 0 & -2.5382  \\
\hline
XVII. $\alpha$  	& -0.1253 & 2.4494 & 0.2506 & -1.6330 & -0.2506  \\
\qquad \enspace  $\beta$       &  0.1053 & 2.2733 & 0.4659 & -1.6032 & -0.4659  \\
\qquad \enspace  $\gamma$      &  0.3656 & 1.9719 & 0.7311 & -1.3146 & -0.7311  \\
\qquad \enspace  $\delta$      & 0.6551 & 1.5440 & 3.1917 &  -2.6796 & -2.7467  \\
\qquad \enspace  $\epsilon$      & 0.8821 & 1.2846 & 2.6931 & -2.1470 & -2.6931  \\
\qquad \enspace  $\zeta$    &  1.0333 & -0.4429 & -0.9614 & 0.7097 & 0.9614 \\
\hline
\end{tabular}
\end{center}
\end{table*}

\begin{table}
\caption{\label{tab:canonsummary} FPs for  $m = 3$ computed in the $\rm\overline{MS}$ scheme to two-loop order. \label{tab4}}
\begin{center}
\begin{tabular}{|c|c|c|c|c|c|}
\hline
FP  & $u^{*}$ & $v^{*}$ & $w^{*}$ & $y^{*}$ & $z^{*}$ \\
\hline
I    & 0 & $ 0 $ & $0$ & $0$ & $0$ \\
\hline
II   &  0 & 1.0016 & 0 & 0 & 0   \\
\hline
III  & 1.5281 & 0 & 0 & 0 & 0  \\
\hline
IV   & 0 & 0 & 1.5281  & 0 & 0  \\
\hline
V    & 0 & 0 & 0 & 1.3146 & 0  \\
\hline
VI   & 0.3411 & 0.6965 & 0 & 0 & 0 \\
\hline
VII  &  2.5382 & 0 & -2.5382 & 0 & 0  \\
\hline
VIII &  0 & 0.8568 & 0 & 0.2270 & 0   \\
\hline
IX. $\alpha$  & -0.2126 & 0.0341 & 0 & 1.5407 & 0 \\
\qquad $\beta$   & 0.3405 & 0.7275 & 0 & -0.0511 & 0 \\
\hline
X. $\alpha$    & 0.7311 & 0 & -0.7311 & 1.3146 & 0   \\
\qquad $\beta$     & -0.0328 & 0 & -0.2134 & 1.6275 & 0 \\
\hline
XI. $\alpha$    & -0.0225 & -0.0003 & -0.2250 & 1.6294 & 0   \\
\qquad $\beta$    & -0.1787 & 0.0175 & -0.0503 & 1.5838 & 0   \\
\qquad $\gamma$     & 0.1822 & -0.0744 & -0.3824 & 1.6437 & 0   \\
\qquad $\delta$    & 0.5908 & 0.4827 & 0.8871 & -0.6083 & 0   \\
\qquad $\epsilon$    &  0.6928 & 1.2393 & 1.3072 & -2.5420 & 0   \\
\hline
XII  & -0.2506 & 0 & 0 & 1.6330 & 0 \\
\hline
XIII & 0 & 0 & -0.2506 & 1.6330 & 0  \\
\hline
XIV  & 0.7126 & 0 & 0 & 0 & 0.5377   \\
\hline
XV   & 1.0728 & 0.7310 & 0 & 0 & -0.5483  \\
\hline
XVI  &  0.7555 & 0 & -0.5382 & 0 & 0.7820    \\
\hline
XVII & 1.1512 & -0.4311 & -1.0150 & 0.9279 & 0.6426 \\
\hline
\end{tabular}
\end{center}
\end{table}

\vspace{-.3cm}
\section{Conclusions}\label{V}
\vspace{-.3cm}

We have studied the effect of generic structural disorder on the  critical properties of  magnets. To that end  we have applied a field-theoretical RG to the RAM with a trimodal distribution of random anisotropy axes which
combines the isotropic and cubic distributions.
We have  derived the RG functions for the model~(\ref{ham_anis}) with arbitrary  $m$ and $n$ to two-loop order. We have used two different regularization schemes, the $\rm\overline{MS}$  scheme and the massive scheme, in order to check the validity of our results.
We have verified that the RG functions reproduce the results known for the limiting cases of the isotropic and cubic distributions.
Applying the Pad\'e-Borel resummation technique we have identified all FPs of the RG flow and
studied their stability. This reveals no stable FP in both schemes except for the FP III, which is unaccessible from physical initial conditions.
This indicates the absence of a continuous phase transition  at variance with the claim  of  Ref.~\onlinecite{Dub17} about the existence of a continuous phase transition of a new universality class. However, as we shown  the conclusion of  Ref.~\onlinecite{Dub17} was based on erroneous two-loop $\beta$ - functions which neither possess the required symmetry properties  nor match with the known results.

Our results show that the magnetic materials with general distribution of random anisotropy axes
do not undergo a continuous phase transition. Although the RG analysis of the type presented here is not able to make a solid conclusion about the origin of the low-temperature phase, our result in combination with other theoretical and numerical data (see in particular the review of results in the introductory part of this paper) gives one more argument in favor of an absence of a low-temperature long-range ordered state.\cite{Dudka05} This is in contrast to the anisotropic distribution of random anisotropy axes where the ferromagnetic order persists in the presence of structural disorder.\cite{Berzin17}
This does not exclude  existence of a QLRO phase similar to that in the case of isotropic distribution of random anisotropies,\cite{feldman04} which, however, is not accessible within our method.

Beside the RAM with a generic random anisotropy distribution,  the RG functions (\ref{gammaALL}), which we have obtained for general $m$ and $n$,  can be also used to study the critical properties of other models
such as  the dilute cubic model\cite{Cal03} and the tetragonal model.\cite{Pelis02}

\vspace{-0.2cm}
\section*{Acknowledgment}
\vspace{-0.2cm}	
Yu. H. and M. D. thank Reinhard Folk and Juan J. Ruiz-Lorenzo for numerous discussions and collaboration. M.D. acknowledges support form Polish National Agency for Academic Exchange through the grant PPN/ULM/2019/00160


\appendix
\numberwithin{equation}{section}
\makeatletter
\newcommand{\section@cntformat}{Appendix \thesection:\ }
\makeatother

\vspace{-0.2cm}
\section{} \label{sec:app0}
\vspace{-0.2cm}

Here we present the relations between our  two-loop  $\beta$-functions computed within $\rm\overline{MS}$ scheme in the limit of $n=0$ ($\beta_{u}, \beta_{v}, \beta_{w}, \beta_{y}, \beta_{z}$) and the $\beta$-functions computed in Ref.~\onlinecite{Korzh88}
for the  phase transition in the crystals with low-symmetry point defects  at replica limit ($\beta_{\lambda}, \beta_{g},  \beta_{\tilde{u}}, \beta_{\tilde{v}}, \beta_{\tilde{w}}$). They read
\begin{widetext}
\begin{eqnarray}
\beta_{ \lambda}(\lambda, g,  \tilde{u}, \tilde{v},  \tilde{w}) &=& {-}\frac{1}{96}\Big[\beta_{y}({-}48 \tilde{v}, 48 g, 48 ( \tilde{v} {+} \tilde{w}{-} \tilde{u}), 48 (\lambda {-} g), {-} 48 \tilde{w}) { +} \beta_{v}({-}48 \tilde{v}, 48 g, 48 ( \tilde{v} {+} \tilde{w}{-} \tilde{u}), 48 (\lambda {-} g), {-} 48 \tilde{w})\Big], \nonumber \\  &&\\
 \beta_{ g}(\lambda, g,  \tilde{u}, \tilde{v},  \tilde{w}) &=& -\frac{1}{96}\beta_{v}({-}48 \tilde{v}, 48 g, 48 ( \tilde{v} {+} \tilde{w}{-} \tilde{u}), 48 (\lambda {-} g), {-} 48 \tilde{w}),  \\
  \beta_{ \tilde{u}}(\lambda, g,  \tilde{u}, \tilde{v},  \tilde{w}) &=& \frac{1}{96} \Big[\beta_{u}({-}48 \tilde{v}, 48 g, 48 ( \tilde{v} {+} \tilde{w}{-} \tilde{u}), 48 (\lambda {-} g), {-} 48 \tilde{w}) + \beta_{w}({-}48 \tilde{v}, 48 g, 48 ( \tilde{v} {+} \tilde{w}{-} \tilde{u}), 48 (\lambda {-} g), {-} 48 \tilde{w}) \nonumber \\ &+& \beta_{z}({-}48 \tilde{v}, 48 g, 48 ( \tilde{v} {+} \tilde{w}{-} \tilde{u}), 48 (\lambda {-} g), {-} 48 \tilde{w})\Big], \\
   \beta_{ \tilde{v}}(\lambda, g,  \tilde{u}, \tilde{v},  \tilde{w}) &=& \frac{1}{96}\beta_{u}({-}48 \tilde{v}, 48 g, 48 ( \tilde{v} {+} \tilde{w}{-} \tilde{u}), 48 (\lambda {-} g), {-} 48 \tilde{w}), \\
    \beta_{ \tilde{w}}(\lambda, g,  \tilde{u}, \tilde{v},  \tilde{w}) &=& \frac{1}{96}\beta_{z}({-}48 \tilde{v}, 48 g, 48 ( \tilde{v} {+} \tilde{w}{-} \tilde{u}), 48 (\lambda {-} g), {-} 48 \tilde{w}).
\end{eqnarray}
\end{widetext}

\section{} \label{sec:secondapp}

In this Appendix we give details on finding the FPs with $z^* \neq 0$ to one-loop order. There are  16 such FPs which can be found along the lines of Ref.~\onlinecite{Korzh88}.  Introducing  variables
\begin{eqnarray}
\label{new_var}
a = u / z, \, \, \, \, \, \, b = v / z, \, \, \, \, \, \, c = w / z, \, \, \, \, \, \, d = y / z,\ \
\end{eqnarray}
we arrive at the system of nonlinear algebraic equations

\begin{subequations}
\label{syst_of_new_var}
\begin{align}
&(m + 4) b^{2} + (6 d + m - n + 6) b + 6 d = 0,  \label{eqa} \\
&(n + 4) c^{2} + (6 d + n - m + 6) c + 6 d = 0, \label{eqb} \\
&9 d^{2} + (8 b + 8 c - m - n + 2) d + 8 b c = 0, \label{eqc} \\
&(4 - m n) a^{2} - (2 m b + 2 n c + 6 d + m + n -2) \nonumber \\
 &- (2 b c + 2 b + 2 c + 3) = 0, \label{eqd} \\
& \varepsilon - z \left(2 a + 2 b / 3 + 2 c /3 + (m + n + 4)/6\right) = 0. \label{eqe}
\end{align}
\end{subequations}
Solving the system of the first three equations (\ref{eqa}) -- (\ref{eqc}) with respect to $b$, $c$, and $d$, we obtain for the case $n\neq0$ eight sets of roots.  Substituting each set ($b, c, d$) into the quadratic equation (\ref{eqd}) we find two values of $a$ leading to $16$ sets ($a, b, c, d$). The corresponding value of $z$ for each set is found from the linear equation (\ref{eqe}). Subsequently,  variables $u, v, w, y$ can be found using (\ref{new_var}). Out of all 16 solutions, only six can be expressed in the analytic form. These are related to three solutions of the system of equations (\ref{eqa}) -- (\ref{eqc}):
\begin{subequations}\label{second:main}
\label{1Looproots}
\begin{align}
b &= c = d = 0, \nonumber \\
b &= - \frac{m - n + 6}{m + 4}, \, c = d = 0, \nonumber \\
b &= 0, \, c = - \frac{n - m + 6}{n + 4}, \ d = 0. \tag{\ref{second:main}}
\end{align}
\end{subequations}
These FPs are denoted by XIV -- XVI in Table~\ref{tab0}. Their coordinates  in the limit $n \to 0$ reproduce the results obtained in Refs.~\onlinecite{Dud01,Korzh88}. The coordinates of the rest $10$ FPs can be found only numerically. Let us note that the  coordinates of these  FPs for the considered values of $m$ and $n$  attain complex values in general. Finally, we have to solve the fifth-order equation for $b$.
The solution of the system of equations (\ref{eqa}) -- (\ref{eqc}) for non-vanishing $n$, reduces to the solution of the following fifth-order equation for $b$:

\begin{flalign}
&(m+4) (3 m-4) (3 m n-4 m-4 n+16) b^{5}&& \nonumber \\ &+ [m^3 (39 n-68)+m^2 \left(-15 n^2+198 n+456\right)
&& \nonumber \\
& +8 m \left(n^2-56 n-72\right)+16 \left(n^2+6 n+24\right)] b^{4}&& \nonumber \\ &+ [m^3 (67 n-148)+m^2 \left(-34 n^2+576 n+1536\right) && \nonumber \\ &+m \left(7 n^3-96 n^2-996 n-2960\right)-4 n^3+96 n^2 && \nonumber \\ &+560 n+1728] b^{3} + [3 m^3 (19 n-52) +m^2 (-19 n^2 && \nonumber \\ &+662 n+2056) -m (5 n^3+88 n^2+1060 n +4112) && \nonumber \\ &-n^4+30 n^3+60 n^2+744 n+2016] b^{2} + 4 [m^3 (6 n && \nonumber \\ &-20)+m^2 \left(n^2+82 n+312\right) -4 m (n^3-5 n^2&& \nonumber \\ &+28 n +164)+n^4-10 n^3+20 n^2 +8 n+352] b&& \nonumber \\ & + 4 (m+n-2) \left(m^2 (n-4)+4 m (5 n+16)-n^3\right.&& \nonumber \\ &\left.-4 n^2-28 n-48\right) = 0.&&
\end{flalign}

Then $c$ and $d$ can be found from:
\begin{eqnarray}
c &=& ( b ((3 m - 4)b + 5 m - n - 2) + 2 (m + n - 2))  \nonumber \\
&& \times \frac{((m + 4)b + m - n + 6)}{6 (b + 1) (b (m - 2) + m - n)}, \\
d &=& - b \frac{(m + 4)b + m - n + 6}{6 (b + 1)}.
\end{eqnarray}
Other parameters can be found using the procedure described above.

The task is simplified in the case $n=0$, since we can extract separate set of roots $b = -2$, $c = - (m + 2)/4$, $d = (m + 2)/3$ in addition to (\ref{1Looproots}).  Therefore we can find the rest 4 roots  solving  the fourth-order equation
\begin{flalign}
&(3 m - 4) (m^{2} - 16) b^{4} + (m-4) (m (11 m+2)-8) b^{3} \nonumber \\ &+ [m (3 m (5 m-36)+196)-112] b^{2}
 \nonumber \\ &+ (m-2) (m (9 m-88)+76) b \nonumber \\ &+ 2 (m-2) ((m-16) m+12) = 0.
\end{flalign}
that can be done analytically.\cite{Abra65}

The rest FPs which can be computed only numerically are shown in Tables~\ref{tabc1}~--~\ref{tabc8}
for several values of $m$ and $n$.
Analysis of these FPs indicates the absence of stable FPs for $m=2$, $m=3$ in the cases $n = 2$ and $n = 3$ (Tables~\ref{tabc5}--\ref{tabc8}). For other values of $n$ ($n = 0$ and $n = 1$) for $m=2$ and $m=3$ only FP III is stable.

\begin{table}
\caption{FPs to the first order in $\varepsilon$ for  $m=2$ and $n=0$. \label{tabc1}}
\begin{center}
\begin{tabular}{|c|c|c|c|c|c|c|}
\hline
FP & $u^{*}$ & $v^{*}$ & $w^{*}$ & $y^{*}$ & $z^{*}$ \\
\hline
I & $0$ & $ 0 $ & $0$ & $0$ & $0$ \\
\hline
II  & $0$ & $ 0.6 $ & $0$ & $0$ & $0$  \\
\hline
III  & $0.75$ & $ 0$ & $0$ & $0$ & $0$  \\
\hline
IV  & $0$ & $ 0 $ & $0.75$ & $0$ & $0$  \\
\hline
V  & $0$ & $ 0 $ & $0$ & $0.6667$ & $0$  \\
\hline
VI  & $-0.75$ & $ 1.5 $ & $0$ & $0$ & $0$  \\
\hline
VII  & $1.5$ & $ 0 $ & $ -1.5 $ & $0$ & $0$  \\
\hline
VIII  & $0$ & $ 1. $ & $0$ & $-0.6667$ & $0$  \\
\hline
X & $0.5$ & $ 0 $ & $-0.5$ & $0.6667$ & $0$  \\
\hline
XI   & $1.5$ & $2.3028$ & $3.4542$ & $-8.1407$ & $0$  \\
     & $0.2295$ & $0.3524$ & $0.5285$ & $-0.3987$ & $0$  \\
     & $0.3631$ & $-0.3153$ & $-0.4730$ & $1.1148$ & $0$  \\
     & $1.5$ & $-1.3028$ & $-1.9542$ & $1.4741$ & $0$  \\
\hline
XIV   & $0.3170$ & $0$ & $0$ & $0$ & $0.3660$ \\
      & $1.1830$ & $0$ & $0$ & $0$ & $-1.3660$ \\
\hline
XV   & $0.2592$ & $-5.7784$ & $0$ & $0$ & $4.3338$ \\
     & $0.5208$ & $0.4984$ & $0$ & $0$ & $-0.3738$ \\
\hline
XVI    & $0.375$ & $0$ & $-0.75$ & $0$ & $0.75$ \\
       & $0.75$ & $0$ & $1.5$ & $0$ & $-1.5$ \\
\hline
XVII    & $0.25$ & $1.$ & $0.5$ & $-0.6667$ & $-0.5$  \\
        & $0.1266$ & $1.3568$ & $0.4730$ & $-1.1148$ & $-0.4730$  \\
        & $3.2262$ & $-9.9050$ & $-3.4530$ & $8.1381$ & $3.4530$  \\
        & $0.5229$ & $0.9083$ & $1.9541$ & $-1.4740$ & $-1.9541$  \\
        & $0.4938$ & $-0.2457$ & $-0.5285$ & $0.3987$ & $0.5285$  \\
\hline
\end{tabular}
\end{center}
\end{table}


\begin{table}
\caption{FPs to the first order in $\varepsilon$ for $m=3$ and $n=0$. \label{tabc2}}
\begin{center}
\begin{tabular}{|c|c|c|c|c|c|c|}
\hline
FP & $u^{*}$ & $v^{*}$ & $w^{*}$ & $y^{*}$ & $z^{*}$ \\
\hline
I   & $0$ & $ 0 $ & $0$ & $0$ & $0$ \\
\hline
II  & $0$ & $ 0.5455 $ & $0$ & $0$ & $0$  \\
\hline
III  & $0.75$ & $ 0$ & $0$ & $0$ & $0$  \\
\hline
IV   & $0$ & $ 0 $ & $0.75$ & $0$ & $0$  \\
\hline
V   & $0$ & $ 0 $ & $0$ & $0.6667$ & $0$  \\
\hline
VI  & $-0.1875$ & $ 0.75 $ & $0$ & $0$ & $0$  \\
\hline
VII  & $1.5$ & $ 0 $ & $-1.5$ & $0$ & $0$  \\
\hline
VIII  & $0$ & $ 0.6667 $ & $0$ & $-0.2222$ & $0$  \\
\hline
IX & $-0.25$ & $ 1. $ & $0$ & $-0.3333$ & $0$  \\
\hline
X & $0.5$ & $ 0 $ & $-0.5$ & $0.6667$ & $0$  \\
\hline
XI   & $0.9313$ & $1.1375$ & $1.9906$ & $-4.2749$ & $0$  \\
     & $0.2411$ & $0.2945$ & $0.5153$ & $-0.3657$ & $0$  \\
     & $0.3691$ & $-0.3453$ & $-0.6043$ & $1.2979$ & $0$  \\
     & $2.8187$ & $-2.6375$ & $-4.6156$ & $3.2749$ & $0$  \\
\hline
XIV   & $0.3158$ & $0$ & $0$ & $0$ & $0.3158$ \\
      & $2.25$ & $0$ & $0$ & $0$ & $-3.$ \\
\hline
XV   & $0.1423$ & $-2.9716$ & $0$ & $0$ & $2.3113$ \\
     & $0.5488$ & $0.4055$ & $0$ & $0$ & $-0.3154$ \\
\hline
XVI    & $0.3462$ & $0$ & $-0.3462$ & $0$ & $0.4615$ \\
       & $1.5$ & $0$ & $2.25$ & $0$ & $-3.$ \\
\hline
XVII    				& $0.3$ & $0.8$ & $0.5$ & $-0.667$ & $-0.4$  \\
                        & $0.2934$ & $0.8369$ & $0.4968$ & $-0.7111$ & $-0.4079$ \\
        				& $0.1644$ & $1.2038$ & $2.8826$ & $-2.6746$ & $-1.7597$ \\
       				    & $0.5678$ & $-0.2432$ & $-0.5823$ & $0.5403$ & $0.3555$ \\
       					& $760.35$ & $-3077.7$ & $-1826.85$ & $2615.25$ & $1500.$  \\
\hline
\end{tabular}
\end{center}
\end{table}


\begin{table}
\caption{FPs to the first order in $\varepsilon$ for $m=2$ and $n=1$. \label{tabc3}}
\begin{center}
\begin{tabular}{|c|c|c|c|c|c|c|}
\hline
FP & $u^{*}$ & $v^{*}$ & $w^{*}$ & $y^{*}$ & $z^{*}$ \\
\hline
I & $0$ & $ 0 $ & $0$ & $0$ & $0$ \\
\hline
II  & $0$ & $ 0.6 $ & $0$ & $0$ & $0$  \\
\hline
III  & $0.6$ & $ 0$ & $0$ & $0$ & $0$  \\
\hline
IV  & $0$ & $ 0 $ & $0.6667$ & $0$ & $0$  \\
\hline
V  & $0$ & $ 0 $ & $0$ & $0.6667$ & $0$  \\
\hline
VI  & $ 3. $ & $ -3. $ & $0$ & $0$ & $0$  \\
\hline
VII  & $1$ & $0$ & $ -0.6667 $ & $0$ & $0$  \\
\hline
VIII  & $0$ & $ 1. $ & $0$ & $-0.6667$ & $0$  \\
\hline
IX  & $1.$ & $ -1. $ & $0$ & $0.6667$ & $0$  \\
\hline
X   & $0.6$ & $ 0 $ & $-0.4$ & $0.4$ & $0$  \\
\hline
XI   & $0.5455$ & $0.4545$ & $0.5455$ & $-1.2121$ & $0$  \\
     & $0.3273$ & $0.2727$ & $0.3273$ & $-0.3273$ & $0$  \\
     & $0.4545$ & $-0.4545$ & $-0.5455$ & $1.2121$ & $0$  \\
     & $1.3636$ & $-1.3636$ & $-1.6364$ & $1.6364$ & $0$  \\
\hline
XII & $1.$ & $ 0 $ & $0$ & $-0.6667$ & $0$  \\
\hline
XIII & $0$ & $0$ & $2.$ & $-2.$ & $0$  \\
\hline
XIV & $0.36$ & $0$ & $0$ & $0$ & $0.24$ \\
      & $1.2$ & $0$ & $0$ & $0$ & $-1.2$ \\
\hline
XV & $0.2308$ & $-1.6154$ & $0$ & $0$ & $1.3846$ \\
     & $0.5538$ & $0.3231$ & $0$ & $0$ & $-0.2769$ \\
\hline
XVI  & $0.3333$ & $0$ & $-0.6667$ & $0$ & $0.6667$ \\
       & $0.6667$ & $0$ & $0.6667$ & $0$ & $-0.6667$ \\
\hline
XVII              & $0.1818$ & $1.3637$ & $0.5455$ & $-1.2122$ & $-0.5455$  \\
				  & $0.8182$ & $-1.3636$ & $-0.5455$ & $1.2121$ & $0.5455$  \\
				  & $0.5454$ & $1.0910$ & $1.6365$ & $-1.6365$ & $-1.6365$ \\
				  & $0.4909$ & $-0.2182$ & $-0.3273$ & $0.3273$ & $0.3273$ \\
				  & $0.4$ & $0.6$ & $0.4$ & $-0.4$ & $-0.4$ \\
				  & $1.$ & $-3.$ & $-2.$ & $2.$ & $2.$ \\
\hline
\end{tabular}
\end{center}
\end{table}


\begin{table}
\caption{FPs to the first order in $\varepsilon$ for $m=3$ and $n=1$. \label{tabc4}}
\begin{center}
\begin{tabular}{|c|c|c|c|c|c|c|}
\hline
FP & $u^{*}$ & $v^{*}$ & $w^{*}$ & $y^{*}$ & $z^{*}$ \\
\hline
I & $0$ & $ 0 $ & $0$ & $0$ & $0$ \\
\hline
II  & $0$ & $ 0.5455 $ & $0$ & $0$ & $0$  \\
\hline
III  & $0.5455$ & $ 0$ & $0$ & $0$ & $0$  \\
\hline
IV  & $0$ & $ 0 $ & $0.6667$ & $0$ & $0$  \\
\hline
V  & $0$ & $ 0 $ & $0$ & $0.6667$ & $0$  \\
\hline
VI  & $6.$ & $ -6. $ & $0$ & $0$ & $0$  \\
\hline
VII  & $0.6667$ & $0$ & $ -0.2222 $ & $0$ & $0$  \\
\hline
VIII  & $0$ & $ 0.6667 $ & $0$ & $-0.2222$ & $0$  \\
\hline
IX & $2.$ & $ -2. $ & $0$ & $0.6667$ & $0$  \\
\hline
X & $0.5455$ & $ 0 $ & $-0.1818$ & $0.1818$ & $0$  \\
\hline
XI & $0.4912$ & $0.1754$ & $0.2456$ & $-0.4678$ & $0$  \\
     & $0.4019$ & $0.1435$ & $0.2010$ & $-0.2010$ & $0$  \\
     & $0.5263$ & $-0.5263$ & $-0.7368$ & $1.4035$ & $0$  \\
     & $1.5790$ & $-1.5790$ & $-2.2105$ & $2.2105$ & $0$  \\
\hline
XII & $0.6667$ & $ 0 $ & $0$ & $-0.2222$ & $0$  \\
\hline
XIII & $0$ & $0$ & $2.$ & $-2.$ & $0$  \\
\hline
XIV & $0.4091$ & $0$ & $0$ & $0$ & $0.1364$ \\
      & $1.5$ & $0$ & $0$ & $0$ & $-1.5$ \\
\hline
XV & $0.1667$ & $-1.3333$ & $0$ & $0$ & $1.1667$ \\
     & $0.5303$ & $0.1212$ & $0$ & $0$ & $-0.1061$ \\
\hline
XVI  & $0.3889$ & $0$ & $-0.2222$ & $0$ & $0.2778$ \\
       & $0.8333$ & $0$ & $0.6667$ & $0$ & $-0.8333$ \\
\hline
XVII            & $0.4091$ & $0.2727$ & $0.2727$ & $-0.2727$ & $-0.1364$  \\
                  & $1.5$ & $-3.$ & $-3.$ & $3.$ & $1.5$  \\
				  & $0.3889$ & $0.5556$ & $0.3333$ & $-0.5556$ & $-0.2778$  \\
				  & $0.8333$ & $-1.6667$ & $-1.$ & $1.6667$ & $0.8333$  \\
				  & $0.1667$ & $0.9998$ & $2.3332$ & $-2.3332$ & $-1.1666$  \\
				  & $0.5303$ & $-0.0909$ & $-0.2121$ & $0.2121$ & $0.1061$  \\
\hline
\end{tabular}
\end{center}
\end{table}


\begin{table}
\caption{FPs to the first order in $\varepsilon$ for $m=2$ and $n=2$. \label{tabc5}}
\begin{center}
\begin{tabular}{|c|c|c|c|c|c|c|}
\hline
FP & $u^{*}$ & $v^{*}$ & $w^{*}$ & $y^{*}$ & $z^{*}$ \\
\hline
I & $0$ & $ 0 $ & $0$ & $0$ & $0$ \\
\hline
II  & $0$ & $ 0.6 $ & $0$ & $0$ & $0$  \\
\hline
III  & $0.5$ & $ 0$ & $0$ & $0$ & $0$  \\
\hline
IV  & $0$ & $ 0 $ & $0.6$ & $0$ & $0$  \\
\hline
V  & $0$ & $ 0 $ & $0$ & $0.6667$ & $0$  \\
\hline
VIII  & $0$ & $ 1. $ & $0$ & $-0.6667$ & $0$  \\
\hline
XIII & $0$ & $0$ & $1.$ & $-0.6667$ & $0$  \\
\hline
XVII              & $0.9$ & $-1.2$ & $-1.2$ & $1.2$ & $0.6$  \\
 				  & $0.3$ & $0.6$ & $1.2$ & $-1.2$ & $-0.6$  \\
				  & $0.3$ & $1.2$ & $0.6$ & $-1.2$ & $-0.6$ \\
				  & $0.5$ & $1.$ & $1.$ & $-1.3333$ & $-1.$ \\
				  & $0.5$ & $-1.$ & $1.$ & $0.6667$ & $1.$ \\
\hline
\end{tabular}
\end{center}
\end{table}


\begin{table}
\caption{FPs to the first order in $\varepsilon$ for $m=3$ and $n=2$. \label{tabc6}}
\begin{center}
\begin{tabular}{|c|c|c|c|c|c|c|}
\hline
FP & $u^{*}$ & $v^{*}$ & $w^{*}$ & $y^{*}$ & $z^{*}$ \\
\hline
I & $0$ & $ 0 $ & $0$ & $0$ & $0$ \\
\hline
II  & $0$ & $ 0.5455 $ & $0$ & $0$ & $0$  \\
\hline
III  & $0.4286$ & $ 0$ & $0$ & $0$ & $0$  \\
\hline
IV  & $0$ & $ 0 $ & $0.6$ & $0$ & $0$  \\
\hline
V  & $0$ & $ 0 $ & $0$ & $0.6667$ & $0$  \\
\hline
VI  & $0.1765$ & $ 0.3529 $ & $0$ & $0$ & $0$  \\
\hline
VII  & $0.2727$ & $0$ & $ 0.2727 $ & $0$ & $0$  \\
\hline
VIII  & $0$ & $ 0.6667 $ & $0$ & $-0.2222$ & $0$  \\
\hline
IX & $0.2$ & $ 0.4 $ & $0$ & $-0.1333$ & $0$  \\
\hline
X & $0.3333$ & $ 0 $ & $0.3333$ & $-0.2222$ & $0$  \\
\hline
XII & $0.3333$ & $ 0 $ & $0$ & $0.2222$ & $0$  \\
\hline
XIII & $0$ & $0$ & $1.$ & $-0.6667$ & $0$  \\
\hline
XV   & $0.2727$ & $-0.5455$ & $0$ & $0$ & $0.5455$ \\
     & $0.3529$ & $-0.3529$ & $0$ & $0$ & $0.3529$ \\
\hline
XVII              & $0.3333$ & $-0.6667$ & $-0.3333$ & $0.2222$ & $0.6667$  \\
   				  & $0.4$ & $-0.4$ & $-0.2$ & $0.1333$ & $0.4$  \\
\hline
\end{tabular}
\end{center}
\end{table}


\begin{table}
\caption{FPs to the first order in $\varepsilon$ for $m=2$ and $n=3$. \label{tabc7}}
\begin{center}
\begin{tabular}{|c|c|c|c|c|c|c|}
\hline
FP & $u^{*}$ & $v^{*}$ & $w^{*}$ & $y^{*}$ & $z^{*}$ \\
\hline
I & $0$ & $ 0 $ & $0$ & $0$ & $0$ \\
\hline
II  & $0$ & $ 0.6 $ & $0$ & $0$ & $0$  \\
\hline
III  & $0.4286$ & $ 0$ & $0$ & $0$ & $0$  \\
\hline
IV  & $0$ & $ 0 $ & $0.5455$ & $0$ & $0$  \\
\hline
V  & $0$ & $ 0 $ & $0$ & $0.6667$ & $0$  \\
\hline
VI  & $0.2727$ & $0.2727$ & $0$ & $0$ & $0$  \\
\hline
VII  & $0.1765$ & $0$ & $0.3529$ & $0$ & $0$  \\
\hline
VIII  & $0$ & $ 1. $ & $0$ & $-0.6667$ & $0$  \\
\hline
IX & $0.3333$ & $0.3333$ & $0$ & $-0.2222$ & $0$  \\
\hline
X & $0.2$ & $ 0 $ & $0.4$ & $-0.1333$ & $0$  \\
\hline
XII & $0.3333$ & $ 0 $ & $0$ & $0.2222$ & $0$  \\
\hline
XIII & $0$ & $0$ & $0.6667$ & $-0.2222$ & $0$  \\
\hline
XVI    & $0.2727$ & $0$ & $-0.5455$ & $0$ & $0.5455$ \\
       & $0.3529$ & $0$ & $-0.3529$ & $0$ & $0.3529$ \\
\hline
XVII    & $0.3333$ & $-0.3333$ & $-0.6667$ & $0.2222$ & $0.6667$  \\
        & $0.4$ & $-0.2$ & $-0.4$ & $0.1333$ & $0.4$  \\
\hline
\end{tabular}
\end{center}
\end{table}


\begin{table}
\caption{FPs to the first order in $\varepsilon$ for $m=3$ and $n=3$. \label{tabc8}}
\begin{center}
\begin{tabular}{|c|c|c|c|c|c|c|}
\hline
FP & $u^{*}$ & $v^{*}$ & $w^{*}$ & $y^{*}$ & $z^{*}$ \\
\hline
I & $0$ & $ 0 $ & $0$ & $0$ & $0$ \\
\hline
II  & $0$ & $ 0.5455 $ & $0$ & $0$ & $0$  \\
\hline
III  & $0.3529$ & $ 0$ & $0$ & $0$ & $0$  \\
\hline
IV  & $0$ & $ 0 $ & $0.5455$ & $0$ & $0$  \\
\hline
V  & $0$ & $ 0 $ & $0$ & $0.6667$ & $0$  \\
\hline
VI  & $0.0896$ & $0.4478$ & $0$ & $0$ & $0$  \\
\hline
VII  & $0.0896$ & $0$ & $0.4478$ & $0$ & $0$  \\
\hline
VIII  & $0$ & $ 0.6667 $ & $0$ & $-0.2222$ & $0$  \\
\hline
IX & $0.1053$ & $0.5263$ & $0$ & $-0.1754$ & $0$  \\
\hline
X & $0.1053$ & $ 0 $ & $0.5263$ & $-0.1754$ & $0$  \\
\hline
\hline
XII & $0.2222$ & $ 0 $ & $0$ & $0.3704$ & $0$  \\
\hline
XIII & $0$ & $0$ & $0.6667$ & $-0.2222$ & $0$  \\
\hline
\end{tabular}
\end{center}
\end{table}

\section{} \label{sec:firstapp}

Here, we present the resummation procedure used in our study. The RG functions calculated within a field-theoretical approach are represented by asymptotic series.
They are characterized by a factorial growth of the coefficients implying a zero radius of convergence.\cite{ZinnJustin96,Amit89} Extracting from them a physical information
requires application of resummation methods, such as the Borel
resummation accompanied by certain additional procedures.\cite{Hardy48}
We use Pad\'e-Borel resummation technique\cite{Bak78} for ``resolvent'' series, where an auxiliary variable is introduced and Borel image of this series is extrapolated by a rational
Pad\'e  approximant [$K/L$] \cite{Bak96} for this new variable.
First, for a given initial polynomial
\begin{equation}
\beta(u, v , w, y, z) = \sum_{1 \leq i + j + k + l + p \leq 5} a_{i, j, k, l, p} u^{i} v^{j} w^{k} y^{l} z^{p},
\end{equation}
we build ``resolvent'' polynomial introducing an auxiliary variable $\lambda$ in the following way:
\begin{eqnarray}
F(u, v, w, y, z; \lambda) &=& \sum_{1 \leq i + j + k + l + p \leq 5} a_{i, j, k, l, p} \nonumber \\
&& \times u^{i} v^{j} w^{k} y^{l} z^{p} \lambda^{i+j+k+l+p-1}, \ \ \  \
\end{eqnarray}
It satisfies  the relation $F(u, v, w, y, z; \lambda {=} 1){ =} \beta(u, v , w, y)$. The Borel image for this series reads
\begin{eqnarray}
\label{serBorel}
F^{B}(u, v, w, y, z; \lambda) &=& \sum_{1 {\leq} i {+} j{ +} k {+} l {+} p {\leq} 5} \frac{a_{i, j, k, l, p} u^{i} v^{j} w^{k} y^{l} z^{p}}{(i{+}j{+}k{+}l{+}p{-}1)!} \nonumber \\
&& \times \lambda^{i+j+k+l+p-1}.
\end{eqnarray}
Subsequently  series (\ref{serBorel}) is approximated by the Pad\'e-approximant $[K/L](\lambda)$, since we are in two-loop approximation we can use only two approximants  $[1/1](\lambda)$, $[0/2](\lambda)$.  It is known that approximants from main diagonal of Pad\'e-matrix \cite{Bak96} have best convergence properties, therefore in our calculations we use $[1/1](\lambda)$ approximant. Finally, the resummed $\beta$-function is found  via inverse Borel transform:
\begin{equation}
\beta^{res}(u, v , w, y, z) = \int_{0}^{\infty} dt \exp(-t)[1/1](t).
\end{equation}

Applying this procedure for the analysis of the RG-functions (\ref{eqbeta})-(\ref{gammaz}) (at the fixed dimension $d=3$) and solving the corresponding system of non-linear FP equations, we obtain the sets of FPs for $m=2$, $m=3$ in the massive scheme as well as the $\rm\overline{MS}$ scheme. Their coordinates are given in Tables~\ref{tab1} -- \ref{tab4}.



\end{document}